\documentclass[12pt,amsmath,amssymb]{revtex4}
\usepackage{amsmath,amsfonts,amsthm,amscd,amssymb,latexsym}
\usepackage{bm}
\usepackage{dcolumn}
\usepackage{graphicx}
\usepackage{epstopdf}
\usepackage{color}
\usepackage{epsf}
\usepackage{epsfig}
\usepackage{graphicx, epic, eepic, color}
\usepackage[colorlinks=true,urlcolor=blue,linkcolor=blue,citecolor=blue]{hyperref}

\begin{document}

\title{Local Peculiar Motions in McVittie and LTB Spacetimes}

\author{Masoud \surname{Molaei}$^{1}$}
\email{masoud.molaei@sharif.ir}
\author{Shant \surname{Baghram}$^{1,2}$}
\email{baghram@sharif.edu}
\author{Bahram \surname{Mashhoon}$^{1,3,4}$} 
\email{mashhoonb@missouri.edu}

\affiliation{$^1$Department of Physics, Sharif University of Technology, Tehran 11155-9161, Iran\\ 
$^2$Research Center for High Energy Physics, Department of Physics, Sharif University of Technology, Tehran 11155-9161, Iran\\ 
$^3$School of Astronomy, Institute for Research in Fundamental Sciences (IPM), Tehran 19395-5531, Iran\\
$^4$Department of Physics and Astronomy, University of Missouri, Columbia,
Missouri 65211, USA\\
}

\date{\today}
\begin{abstract}
We consider two inhomogeneous cosmological models, namely, the flat McVittie spacetime and a simple specific LTB spacetime. Relative to the world line of a reference comoving observer that remains spatially at rest, we study the local deviations of the world lines of free test particles. These local peculiar motions can be invariantly characterized within the framework of a quasi-inertial Fermi normal coordinate system established along the world line of the reference comoving observer. Tidal dynamics in the McVittie model involves the sum of the curvature due to the inhomogeneity, the curvature due to the background FLRW spacetime and a mixed term, while tidal dynamics in the particular LTB model turns out to be qualitatively the same as in the Einstein-de Sitter universe. Peculiar motions in the two cosmological models are briefly compared and contrasted.
\end{abstract}

\maketitle

% @@@@@@@@@@@@@@@@@@@@@
% @@@@@@@@@@@@@@@@@@@@@
% @@@@@@@@@@@@@@@@@@@@@

\section{Introduction}

A free test particle follows a geodesic in a gravitational field. If the background field is stationary, there is a timelike Killing vector field and the projection of the 4-velocity of the test particle on this Killing vector is a constant of the motion. We interpret this circumstance as indicating that the  test particle along its world line does not exchange energy with the background field and that the energy of the particle is thus conserved as a consequence of the invariance of the gravitational field under translation in time. The situation is different, however, if, as in cosmology, the gravitational field is time dependent~\cite{Bini:2014esa, Chicone:2010xr, Chicone:2011ie}. 

In the standard Friedmann-Lema\^itre-Robertson-Walker (FLRW) cosmology, peculiar motions refer to the deviations of free particles from the Hubble flow; moreover, peculiar motions are usually treated within the normal framework of cosmological perturbation theory. For instance, the dispersion of velocities in clusters of galaxies  typically amounts to $v_{pec}/c \sim 10^{-3}$~\cite{Dodelson:2003ft, Weinberg:2008zzc, Amendola:2015ksp, Baumann:2022mni}. 

The observational data contained in the cosmic web indicates the presence of inhomogeneity at various scales in the distribution of structure in the universe; therefore, large-scale peculiar motions must exist due to the gravitational attraction of mass-energy. This notion is supported by ample observational evidence for peculiar motions; see, for instance,~\cite{Mohayaee:2020wxf, Immer, Zin, Pesce, Giahi-Saravani:2012rou, Baba:2009ep} and the references therein. 

To develop a fully relativistic theory of geodesic motion relative to the class of preferred comoving observers that are spatially at rest in a cosmological model, we study the motion of free particles in a quasi-inertial Fermi normal coordinate system established along the world line of a fiducial preferred comoving observer. In this approach, the state of the preferred observers becomes rather significant. In the standard FLRW cosmological model, for instance, the energy density and pressure only depend upon time and the preferred comoving observers that constitute the Hubble flow thus follow spacetime geodesics. This is not always the case in the inhomogeneous cosmological models that are the focus of the present work.  

Imagine a cosmological model with a spacetime metric written in comoving coordinates $x^\mu$ as
\begin{equation}\label{1}
ds^2 = g_{\mu \nu}(x)\, dx^\mu \,dx^\nu\,, 
\end{equation}
which satisfies Einstein's field equations~\cite{Einstein}
\begin{equation}\label{2}
G_{\mu \nu} + \Lambda\, g_{\mu \nu}=\kappa\,T_{\mu \nu}\,, 
 \end{equation}
where $T_{\mu \nu}$ is the symmetric energy-momentum tensor of matter, $\kappa:=8 \pi G/c^4$, $\Lambda$ is the cosmological constant and $G_{\mu \nu}$ is the Einstein tensor  
\begin{equation}\label{3}
G_{\mu \nu} := R_{\mu \nu}-\frac{1}{2} g_{\mu \nu}\,g^{\alpha \beta}R_{\alpha \beta}\,.
 \end{equation} 
Here, Greek indices run from 0 to 3, while Latin indices run from 1 to 3; moreover, the signature of the metric is +2 and we use natural units such that $c = G = 1$, unless specified otherwise. 

It is usually assumed that the cosmological source can be approximated by a perfect fluid of the form   
\begin{equation}\label{4}
T^{\mu \nu} =  (\mu + p ) u^\mu u^\nu + p g^{\mu \nu}\,, 
\end{equation}
where $\mu$ is the energy density, $p$ is the pressure and $u^\mu = dx^\mu /d\tau$  is the 4-velocity vector of the perfect fluid that is comoving, namely,  it is spatially at rest (i.e., $u^\mu = 0$ for $\mu \ne 0$). Here, $\tau$ is the proper time of the perfect fluid.  The proper time $\tau$ is shared by the preferred observers that are comoving with the perfect fluid. 
 It follows from $T^{\mu \nu}{}_{; \nu} = 0$ that 
\begin{equation}\label{5}
(\mu + p ) u^{\nu}{}_{; \nu}  = \frac{d\mu}{d\tau}\,, \qquad (\mu + p ) \frac{Du^\mu}{d\tau} = -(g^{\mu \nu} + u^\mu u^\nu) \frac{\partial p}{\partial x^\nu}\,. 
\end{equation}
Under reasonable conditions, $\mu + p  \ne 0$ and the Hubble flow is geodesic if the transverse (i.e., spatial) gradient of the pressure vanishes. This is the case in the homogeneous FLRW model, where the pressure only depends upon time, as well as in the 
inhomogeneous Szekeres~\cite{Szekeres:1974ct} and Lema\^itre-Tolman-Bondi~\cite{Lemaitre:1933gd,Tolman:1934za, Bondi:1947fta} (LTB) dust models that have vanishing pressure and are particular generalizations of the FLRW model~\cite{Krasinski:1997yxj, Plebanski:2006sd, Bolejko:2009pvd}. For further work on inhomogeneous cosmology see, for instance,~\cite{Ribeiro:1992iwc, Marra:2007pm, Marra:2007gc, Duffy:2010bu, Cosmai:2018nvx} and the references therein. 

On the other hand, if the spatial gradients of the pressure do not all vanish, the Hubble flow is not geodesic. An example is the inhomogeneous McVittie model~\cite{McVittie:1933zz}. The McVittie universe involves the embedding of a point mass $M$ in the FLRW universe. As is well known, the metric of the FLRW universe depends on a parameter $k = 1$, $-1$, or $0$, for the closed, open, or flat model, respectively. In these three cases, the perfect fluid source of the McVittie spacetime has pressure with nonzero spatial gradients resulting in the acceleration of test observers that are spatially at rest. This outward acceleration prevents the radial infall of cosmic matter onto the inhomogeneity. The divergence of the pressure in the McVittie spacetime indicates the presence of a curvature singularity. In contrast, the spacetime singularities in the LTB dust models occur where the energy density diverges~\cite{Krasinski:1997yxj, Plebanski:2006sd, Bolejko:2009pvd}. 

The McVittie model is interesting as it seems to represent a black hole embedded in the FLRW universe. On the other hand, the McVittie model exhibits some rather odd features. We are interested in the local peculiar motions in the flat McVittie model. To place our work in a broader context, we study peculiar motions in another inhomogeneous cosmological model and note the similarities and differences between the two models. 

Local peculiar motions are important within both the cosmological as well as the astrophysical contexts. On cosmological scales, the velocity field of matter distribution on large scales is a probe to measure the growth of structures via redshift space distortion~\cite{Amendola:2015ksp,  Hahn:2014lca} or to study the bulk flow, which is the average of peculiar velocities in smoothed spheres~\cite{Watkins:2008hf}.  On astrophysical scales, studying the peculiar velocities of the hosts of the supernovas is essential for their usage as standard candles~\cite{Davis:2010jq}. Accordingly, investigating the McVitte model is a step forward in addressing these measurements in more realistic models.

The main purpose of this paper is therefore to compare and contrast peculiar motions in two different yet somewhat similar inhomogeneous cosmological models, namely, the flat $(k = 0)$ McVittie model and a simple LTB model. We study motions relative to a reference preferred comoving observer within the framework of a quasi-inertial Fermi normal coordinate system established along the world line of the fiducial comoving observer~\cite{Synge, mash77, Chicone:2002kb, Chicone:2005vn}. Fermi coordinates have been employed in the cosmological context before; see, for instance~\cite{Mashhoon, Cooperstock:1998ny,Mashhoon:2007qm}. For other useful approaches to the equations of motion in cosmology see, for instance,~\cite{Nandra:2011ug, Nandra:2011ui, Nandra:2013jga} and the references cited therein. 

The plan of this paper is as follows. A brief description of the flat McVittie model is provided in Section II. In Section III, we establish a quasi-inertial Fermi normal coordinate system along the local Fermi-Walker transported tetrad frame of the fiducial preferred comoving observer and in Section IV discuss geodesic (i.e., peculiar) motion in the Fermi coordinate system. The peculiar motions in the flat McVittie case are compared and contrasted with peculiar motions in a simple LTB model in Section V. Section VI contains a discussion of our results. 

\section{Flat McVittie Model}

In 1933, McVittie~\cite{McVittie:1933zz} published an exact solution of Einstein's field equations of general relativity that was a nonlinear superposition of the Schwarzschild spacetime and the FLRW cosmological model and represented a point mass in an expanding universe. The original motivation for this kind of study was to determine the influence of the expansion of the universe on local physics; indeed, a considerable body of literature exists on this problem, see~\cite{Cooperstock:1998ny, Mashhoon:2007qm, Faraoni:2007es, Kopeikin:2012by, Kopeikin:2013am, Iorio:2012wva, Spengler:2021vxy} and the references therein.   The McVittie spacetime has been the subject of detailed investigations; see~\cite{Ferraris:1996ey, Bonnor, Sakai:1999xx,  Kaloper:2010ec, Lake:2011ni, Nolan:2014maa, Nolan:2017rtj, Perlick:2018iye, Faraoni:2018xwo, Rothman:2018haq, Gaur:2023hmk} and the references cited therein.

We are interested in the McVittie solution only for the case where the FLRW universe is spatially flat.The flat McVittie metric, 
\begin{equation}\label{I1}
 ds^2 = - \left[\frac{1-\frac{M}{2\rho\,a(t)}}{1+\frac{M}{2\rho\,a(t)}}\right]^2 \,dt^2 + a^2(t)\,[1+\tfrac{M}{2\rho\,a(t)}]^4\,\delta_{ij}\,dx^i\,dx^j\,, 
\end{equation} 
is a nonlinear superposition of the isotropic Schwarzschild solution and the flat FLRW model~\cite{McVittie:1933zz}. Here, $\rho = (\delta_{ij} x^i x^j)^{1/2}$ is the radial coordinate and $a(t)$ is the scale factor.  The two limits are obtained for $a(t) =1$ and $M = 0$, respectively. 

Metric~\eqref{I1}  satisfies the gravitational field equations of general relativity~\eqref{2} with $\Lambda = 0$ and a perfect-fluid $T_{\mu \nu}$ given by 
\begin{equation}\label{I2}
T_{\mu \nu}=\mu_{\rm Mc}\,u_\mu\,u_\nu+p_{\rm Mc}\,(g_{\mu \nu}+u_\mu\,u_\nu)\,.
\end{equation}
Here, $\mu_{\rm Mc}$, $p_{\rm Mc}$ and $u^\mu$ are the energy density, pressure and the 4-velocity vector of the perfect fluid, respectively. As in the standard cosmological models, we assume the fluid particles are spatially at rest and comoving with the preferred observers. That is, 
\begin{equation}\label{I3}
 u^\mu = \frac{dx^\mu}{d\tau} = \frac{Q}{P} \,\delta^\mu_0\,, \qquad u^\mu u_\mu = -1\,,
\end{equation}
where $\tau$ is the proper time and
\begin{equation}\label{I4}
 P = 2 \rho a - M\,, \qquad Q = 2 \rho a + M\,.
\end{equation}  
The motion of the fluid is shear free. The density and pressure of the flat McVittie solution are given by
\begin{equation}\label{I5}
\mu_{\rm Mc} = \frac{3H^2}{8 \pi}\,, \qquad p_{\rm Mc} = - \frac{3H^2}{8 \pi} -\frac {1}{4\pi} \,\frac{Q}{P}\, \dot{H}\,,
\end{equation}
where $H$ is the Hubble parameter ($\dot{a}/a$), namely, 
\begin{equation}\label{I6}
H := \frac{1}{a} \frac{da}{dt}\,, \qquad \dot{H} := \frac{dH}{dt}\,.
\end{equation}  
The density is uniform, while the pressure is nonuniform and diverges at $P = 0$ for $\dot{H} \ne 0$. We show below that the flat McVittie solution reduces to the Schwarzschild-de Sitter solution if the Hubble parameter is constant in time. Therefore, let us suppose that $\dot{H} \ne 0$; then, the big bang singularity occurs in the flat McVittie solution where $2 \rho\, a(t) = M$.  This is a spacelike hypersurface where $\sqrt{-g} = 0$ for metric~\eqref{I1}.

For $\rho \gg M/a(t)$, the McVittie overdensity slowly disappears and the flat McVittie model approaches the flat FLRW model with $\mu_{\rm FLRW} (k = 0) = \mu_{\rm Mc}$ and $p_{\rm FLRW} (k = 0) = -\tfrac{3}{8 \pi} H^2 -\tfrac{1}{4 \pi} \dot{H}$. 

It proves useful to introduce a new radial coordinate $\rho'$ given by
\begin{equation}\label{I7}
 \rho' := \rho\, a(t)\,, \qquad  a(t)\, d\rho = d\rho' - H \rho' dt\,, 
\end{equation} 
where $H$ is the Hubble parameter given by Eq.~\eqref{I6}. In spherical polar coordinates $(\rho', \theta, \phi)$, the McVittie metric now takes the form
\begin{equation}\label{I8}
 ds^2 = - \left(\frac{1-\frac{M}{2\rho'}}{1+\frac{M}{2\rho'}}\right)^2 \,dt^2 + (1+\tfrac{M}{2\rho'})^4\,[(d\rho' - H \rho' dt)^2 + \rho'^2 d\Omega^2]\,, 
\end{equation} 
where
\begin{equation}\label{I9}
d\Omega^2 = d\theta^2 + \sin^2 \theta \,d\phi^2\,.
\end{equation} 
Next, let us introduce the Schwarzschild-like radial coordinate $\mathfrak{r}$ given by
\begin{equation}\label{I10}
\mathfrak{r} := \rho' (1+\tfrac{M}{2\rho'})^2\,.  
\end{equation} 
This definition then implies
\begin{equation}\label{I10a}
\mathfrak{r} - 2\,M = \rho' (1-\tfrac{M}{2\rho'})^2\,,\qquad \frac{d\rho'}{\rho'} = \frac{d\mathfrak{r}}{\mathfrak{r}\,\sqrt{1-\tfrac{2M}{\mathfrak{r}}}}\,. 
\end{equation} 
The flat McVittie metric  can now be written as 
\begin{equation}\label{I11}
 ds^2 = - (1-\tfrac{2M}{\mathfrak{r}} - H^2 \mathfrak{r}^2) dt^2   -2 \frac{H\,\mathfrak{r}}{\sqrt{1-\tfrac{2M}{\mathfrak{r}}}}\,dt\,d\mathfrak{r} + \frac{d\mathfrak{r}^2}{1-\tfrac{2M}{\mathfrak{r}}} + \mathfrak{r}^2  d\Omega^2\,.
\end{equation}
These $(t, \mathfrak{r}, \theta, \phi)$ coordinates are admissible for $H^2\mathfrak{r}^2< 1 -2\,M/\mathfrak{r}$; moreover, for metric~\eqref{I11}, $\sqrt{-g} = \mathfrak{r}^2 \sin\theta$. 

The inverse metric has nonzero components
\begin{equation}\label{I12}
 g^{tt} = -\frac{1}{1-\tfrac{2M}{\mathfrak{r}}}\,, \quad g^{t\mathfrak{r}}= g^{\mathfrak{r}t} =  - \frac{H\,\mathfrak{r}}{\sqrt{1-\tfrac{2M}{\mathfrak{r}}}}\,, 
\end{equation}
\begin{equation}\label{I13}
 g^{\mathfrak{r}\mathfrak{r}} = (1-\tfrac{2M}{\mathfrak{r}} - H^2 \mathfrak{r}^2)\,, \quad g^{\theta \theta}= \frac{1}{\mathfrak{r}^2}\,, \quad g^{\phi \phi}= \frac{1}{ \mathfrak{r}^2\sin^2\theta}\,.
\end{equation}

Let us note that in the weak-field approximation, we can write $-g_{00} = 1 + 2 \Phi_{\rm N}$, where the Newtonian potential is $\Phi_{\rm N} = - \tfrac{M}{\mathfrak{r}} - \tfrac{1}{2} H^2 \mathfrak{r}^2$. In this limiting situation, the Newtonian potential at $\mathfrak{r}$ is consistent within the context of Newtonian gravitation with a point mass $M$ located at $\mathfrak{r}= 0$ and a uniform distribution of perfect fluid of density $3H^2/(8\,\pi)$ occupying a sphere of radius $\mathfrak{r}$. 

If $H = {\rm constant}$, as in the de Sitter case ($H^2 = \Lambda/3$), the transformation $t \mapsto \bar{t}$, where 
\begin{equation}\label{I14}
dt = d\bar{t} - \frac{H\mathfrak{r}}{(1-\tfrac{2M}{\mathfrak{r}} - H^2 \mathfrak{r}^2)}\,\frac{d\mathfrak{r}}{\sqrt{1-\tfrac{2M}{\mathfrak{r}}}}\,, 
\end{equation} 
transforms the McVittie metric to the Schwarzschild-de Sitter metric 
\begin{equation}\label{I15}
 ds^2 = - (1-\tfrac{2M}{\mathfrak{r}} - H^2 \mathfrak{r}^2) d\bar{t}^2  + \frac{d\mathfrak{r}^2}{(1-\tfrac{2M}{\mathfrak{r}} - H^2 \mathfrak{r}^2)} + \mathfrak{r}^2  d\Omega^2\,
\end{equation} 
with $H^2 = \Lambda/3$. The Schwarzschild-de Sitter solution of general relativity was originally found by Kottler~\cite{Kottler}.

It would be interesting to work out the Kretschmann scalar $\mathcal{K} = R_{\alpha\beta\gamma\delta} R^{\alpha\beta\gamma\delta}$ for metric~\eqref{I11} and the result is
\begin{equation}\label{I16}
\frac{1}{12}\, \mathcal{K}=  \frac{2 H^2(t) \dot{H}(t)}{\sqrt{1-\frac{2 M }{\mathfrak{r}}}}+\frac{\mathfrak{r} \dot{H}^2(t)}{\mathfrak{r}-2 M }+2 H^4(t)+\frac{4 M ^2}{\mathfrak{r}^6}\,.
\end{equation}
If the Hubble parameter is not a constant, then the spacetime singularity occurs at $\mathfrak{r}=2\,M$; otherwise, the flat McVittie metric reduces to the Schwarzschild-de Sitter metric which is singular at the black hole singularity $\mathfrak{r}=0$; that is, 
 the Kretschmann scalar for the Kottler spacetime is given by
\begin{equation}\label{I17}
\mathcal{K}|_{\rm Kottler} =  \frac{8}{3} \Lambda^2 + 48\,\frac{M ^2}{\mathfrak{r}^6}\,.
\end{equation}
Moreover,   
the scalar curvature for metric~\eqref{I11} is given by
\begin{equation}\label{I18}
\frac{1}{6}\,g^{\mu \nu}R_{\mu \nu} =\frac{ \dot{H}(t)}{\sqrt{1-\frac{2 M }{\mathfrak{r}}}}+2 H^2(t)\,.
\end{equation}

Let us return to the flat McVittie metric~\eqref{I1} and look at it within the context of spherically symmetric spacetimes. A general spherically symmetric spacetime in comoving coordinates $x^\mu = (t, \rho, \theta, \phi)$ has a metric of the form
\begin{equation}\label{I19}
ds^2 = - \mathbb{A}^2(t, \rho)\, dt^2 + \mathbb{B}^2(t, \rho)\, d\rho^2 + \mathbb{R}^2(t, \rho)\,(d\theta^2 + \sin^2 \theta \,d\phi^2)\,,
\end{equation}
where $\mathbb{A}$, $\mathbb{B}$ and $\mathbb{R}$ are functions of time $t$ and radial coordinate $\rho$. In the general relativistic dynamics of these gravitational fields, the amount of mass-energy $m(t, \rho)$ within a radius $\rho$ at time $t$ plays a significant role. This quantity, known as the Misner-Sharp mass~\cite{Misner:1964je, HeMi, CaMc}, is an invariant such that $2\,m(t, \rho)/\mathbb{R}^3(t, \rho)$ is the sectional curvature in the direction of the surface of the sphere, and it can be expressed in terms of the gravitational potentials as~\cite{GlaMa, Mashhoon:1979tt}
\begin{equation}\label{I20}
m(t, \rho) = \frac{1}{2} \mathbb{R} \left[ 1 + \left(\frac{1}{\mathbb{A}}\frac{\partial \mathbb{R}}{\partial t}\right)^2 -  \left(\frac{1}{\mathbb{B}}\frac{\partial \mathbb{R}}{\partial \rho}\right)^2\right]\,.
\end{equation}

To find the Misner-Sharp mass-energy function for the flat McVittie spacetime, let us write Eq.~\eqref{I1} in a similar form as Eq.~\eqref{I19}, namely, 
\begin{equation}\label{J1}
ds^2 = - \frac{P^2}{Q^2}\,dt^2 + \frac{Q^4}{16\,\rho^4\,a^2}\,(d\rho^2 + \rho^2d\theta^2 + \rho^2 \sin^2\theta d\phi^2)\,,
\end{equation} 
where $(P, Q) = (2\rho a - M, 2\rho a + M)$. A straightforward calculation using Eq.~\eqref{I20} reveals that 
\begin{equation}\label{J2}
m_{\rm Mc}(t, \rho) = M + \frac{1}{2} (\rho a)^3 H^2 \left(1 +\frac{M}{2\rho a}\right)^6\,.
\end{equation} 
Let us note that this is the mass-energy at time $t$ within a sphere of radius $\rho$ beyond the singularity; that is, $2 \rho a > M$. It follows from Eq.~\eqref{J2} that the Schwarzschild mass $M$ is not influenced by the expansion of the McVittie universe; however, it does have an effect on the mass-energy content of the flat FLRW background; indeed, for $M = 0$, the second term in  $m_{\rm Mc}$ is given by
\begin{equation}\label{J3}
 \frac{4 \pi}{3} (\rho a)^3\, \frac{3H^2}{8 \pi}\,,
\end{equation} 
where $ \mu_{\rm Mc} = 3H^2/(8 \pi)$ is the energy density of the flat FLRW model as well as of the McVittie universe by Eq.~\eqref{I5}. 

The Christoffel symbols for metric~\eqref{J1} are given in Appendix A. Finally, we note that the scalar curvature and the Kretschmann scalar for metric~\eqref{J1} are given by
\begin{equation}\label{J4}
g^{\mu \nu} R_{\mu \nu} = 12 H^2 + 6 \dot{H}\frac{Q}{P}\,,
\end{equation} 
\begin{equation}\label{J5}
R_{\alpha\beta\gamma\delta} R^{\alpha\beta\gamma\delta} = 12\left(2\, H^4+2\,\frac{Q}{P}\, H^2 \dot{H}+ \frac{Q^2}{P^2}\, \dot{H}^2 +16384\, \frac{M^2a^6\rho^6}{Q^{12}}\right)\,. 
\end{equation} 

\section{Fermi Normal Coordinates for Flat McVittie Spacetime}

We wish to concentrate here on the measurements of the class of preferred observers that are spatially at rest in this spacetime with 4-velocity vector $u^\mu$ given by Eq.~\eqref{I3}.  In $(t, \rho, \theta, \phi)$ coordinates, let us introduce the orthonormal tetrad frame $e^{\mu}{}_{\hat \alpha}$,
\begin{equation}\label{F1}
e^{\mu}{}_{\hat 0} := u^\mu = \frac{Q}{P} \,\delta^\mu_0\,,  \quad e^{\mu}{}_{\hat 1} := \frac{4 a \rho^2}{Q^2}\, \delta^\mu_{1}\,,\quad e^{\mu}{}_{\hat 2} := \frac{4 a \rho}{Q^2}\, \delta^\mu_{2}\,, \quad e^{\mu}{}_{\hat 3} := \frac{4 a \rho}{Q^2\sin\theta}\, \delta^\mu_{3}\,,
\end{equation} 
which is adapted to the preferred observers. The local spatial frame of a comoving observer has unit axes that point along the spatial coordinate directions. It follows from the orthonormality relation
\begin{equation}\label{F2}
g_{\mu \nu}\,e^{\mu}{}_{\hat \alpha}\,e^{\nu}{}_{\hat \beta} = \eta_{\hat \alpha \hat \beta}\,
\end{equation} 
that the acceleration tensor $\Phi_{\hat \alpha \hat \beta}$ of these observers  given by
\begin{equation}\label{F3}
\frac{D e^{\mu}{}_{\hat \alpha}}{d\tau} = \Phi_{\hat \alpha}{}^{\hat \beta} \, e^{\mu}{}_{\hat \beta}\,
\end{equation} 
is antisymmetric, namely, $\Phi_{\hat \alpha \hat \beta} = - \Phi_{\hat \beta \hat \alpha}$. Using the connection coefficients given in Appendix A, we find that the only nonzero components of the acceleration tensor are
\begin{equation}\label{F4}
\Phi_{\hat 0 \hat 1} = - \Phi_{\hat 1 \hat 0} = \gamma = 16Ma^2\rho^2/(PQ^3)\,.
\end{equation} 
 Note that $\gamma$ is positive, which means that for the observer to stay fixed in space, this quantity balances the attraction of gravity. In this connection, $\gamma \to \infty$ at the spacetime singularity where $\rho \to M/(2a)$, while $\gamma \to M/(a^2\rho^2)$ for $\rho \to \infty$, which is the Newtonian result and indicates that in this limit $a(t)\,\rho$ is the appropriate Newtonian radial distance. 

To gain insight into the nature of McVittie's gravitational field, it is useful to establish local quasi-inertial Fermi normal coordinate systems~\cite{Synge, mash77, Chicone:2002kb, Chicone:2005vn} in the neighborhoods of preferred observers. For the flat  McVittie universe, we choose the congruence of  observers that are spatially at rest and carry the natural orthonormal tetrad system~\eqref{F1}. We choose one such observer with fixed spatial coordinates $(\bar{\rho}, \bar{\theta}, \bar{\phi})$ to be the reference observer that stays away from the spacetime singularity, namely, $2\bar{\rho} > M/a(t)$. In this connection, it proves useful to define $\alpha > 0$ such that
\begin{equation}\label{N1}
\alpha + 1 := \frac{\bar{Q}}{\bar{P}} = \frac{2 \bar{\rho} a(t)+ M}{2 \bar{\rho} a(t)- M} > 1\,
\end{equation}
and the proper time of the reference observer is then given by
\begin{equation}\label{N2}
\tau = c\,\int \frac{dt}{1 + \alpha(t)}\,.
\end{equation}
 
Let $\bar{x}^\mu(\tau) = ( t, \bar{\rho}, \bar{\theta}, \bar{\phi})$ be the world line of the fiducial observer and $\bar{e}^{\mu}{}_{\hat \alpha}$ be the corresponding adapted tetrad frame. At an arbitrary event with proper time $\tau$ along the world line of the reference observer, consider the class of all spacelike geodesics that originate normally from this event and generate a local spacelike hypersurface. Let $x^\mu$ be an event on this hypersurface that can be connected to $\bar{x}^\mu(\tau)$ by a unique spacelike geodesic with proper length $\sigma$; then, to event $x^\mu$ we assign invariantly defined Fermi coordinates $X^{\hat \mu}$ such that   
\begin{equation}\label{N3}
X^{\hat 0} := \tau\,, \qquad X^{\hat i} := \sigma\, \xi^\mu(\tau)\, \bar{e}_{\mu}{}^{\hat i}(\tau)\,,
\end{equation} 
where $\xi^\mu(\tau)$ is the unit spacelike vector tangent to the unique  geodesic segment of proper length $\sigma$ at $\bar{x}^\mu(\tau)$.  The reference observer in the flat McVittie universe has translational acceleration  given by Eq.~\eqref{F4} and its spatial frame is Fermi-Walker transported along its world line. In this case, the spacetime metric in Fermi coordinates is given by
\begin{equation}\label{N4}
ds^2 = g_{\hat \mu \hat \nu}\,dX^{\hat \mu} dX^{\hat \nu}\,,
\end{equation}
where
\begin{equation}\label{N5}
g_{\hat 0 \hat 0} = - (1+\gamma X^{\hat 1})^2  - R_{\hat 0 \hat i \hat 0 \hat j}\,X^{\hat i}\,X^{\hat j}\,,
\end{equation}
\begin{equation}\label{N6}
g_{\hat 0 \hat i} = -\frac{2}{3} \,R_{\hat 0 \hat j \hat i \hat k}\,X^{\hat j}\,X^{\hat k}\,
\end{equation}
and
\begin{equation}\label{N7}
g_{\hat i \hat j} = \delta_{\hat i \hat j} -\frac{1}{3} \,R_{\hat i \hat k \hat j \hat l}\,X^{\hat k}\,X^{\hat l}\,,
\end{equation}
where  third and higher-order terms in spatial Fermi coordinates have been neglected for the sake of simplicity.  
The Fermi coordinate system is admissible in a sufficiently narrow cylindrical spacetime domain along the reference world line; in fact, the spatial Fermi coordinates should be sufficiently small compared to the local radius of curvature of spacetime~\cite{Chicone:2005vn}.

In the case of the flat McVittie spacetime under consideration here, we note that
\begin{equation}\label{N8}
R_{\hat \alpha \hat \beta \hat \gamma \hat \delta} = R_{\mu \nu \rho \sigma}\,e^{\mu}{}_{\hat \alpha}\,e^{\nu}{}_{\hat \beta}\,e^{\rho}{}_{\hat \gamma}\,e^{\sigma}{}_{\hat \delta}\,
\end{equation}
is the projection of the Riemann curvature tensor upon the fiducial observer's tetrad frame.
In general, we can express Eq.~\eqref{N8} as a $6\times 6$ matrix with indices that range over the set $\{01,02,03,23,31,12\}$. In an arbitrary gravitational field, we find
\begin{equation}\label{N9}
\left[
\begin{array}{cc}
\mathcal {E} & \mathcal{B}\cr
\mathcal{B}^{\rm T} & \mathcal{S}\cr 
\end{array}\right]\,,
\end{equation}
where $\mathcal{E}$ and $\mathcal{S}$ are symmetric $3\times 3$ matrices and $\mathcal{B}$ is traceless due to the symmetries of the Riemann tensor. In a Ricci-flat spacetime, $\mathcal{E}$ and $\mathcal{B}$ are symmetric and traceless, while $\mathcal{S} = - \mathcal{E}$. Here,  $\mathcal{E}$, $\mathcal{B}$ and $\mathcal{S}$ denote the gravitoelectric,  gravitomagnetic and spatial components of the Riemann curvature tensor  as measured by the reference observer, respectively.  In the flat McVittie universe, we find 
\begin{equation}\label{N10}
\mathcal{E} = {\rm diag}(\mathcal{E}_1, \mathcal{E}_2, \mathcal{E}_3)\,, \qquad \mathcal{B} = 0\,,\qquad \mathcal{S} = {\rm diag}(\mathcal{S}_1, \mathcal{S}_2, \mathcal{S}_3)\,,
\end{equation} 
where
\begin{equation}\label{N11}
\mathcal{E}_1 + \mathcal{S}_1= \mathcal{E}_2 + \mathcal{S}_2 = \mathcal{E}_3 + \mathcal{S}_3 = - (\alpha + 1)\,\dot{H}\,,
\end{equation}
\begin{equation}\label{N12}
\mathcal{S}_1=  H^2 + 2 \beta\,, \qquad \mathcal{S}_2 = \mathcal{S}_3 = H^2 - \beta\,.
\end{equation}
Let us note that  $-\dot{H} - H^2 = q H^2$ implies
\begin{equation}\label{N12a}
\mathcal{E}_1 =  - 2\,\beta + qH^2 -\alpha \dot{H}\,, \qquad \mathcal{E}_2 = \mathcal{E}_3 = \beta + qH^2 -\alpha \dot{H}\,, 
\end{equation}
where we can use $\rho' = \rho\, a(t)$ defined in Eq.~\eqref{I7} to write
\begin{equation}\label{N12b}
\alpha = \frac{2\,M}{2\rho'-M} = \frac{M}{\rho'} \,\left(1- \frac{M}{2\,\rho'}\right)^{-1}\,,
\end{equation}
\begin{equation}\label{N13}
\beta := \frac{64 \,M  \rho ^3 \mathit{a}^3(t)}{Q^6} = M\,\left[\rho'\,\left(1+ \frac{M}{2\,\rho'}\right)^2\right]^{-3}\,.
\end{equation}
All of these quantities are functions of $T:=X^{\hat 0}$ and are evaluated along the world line of the reference observer. The curvature of the flat McVittie spacetime as measured by comoving observers has an interesting structure, namely, it is the sum of the curvature of the inhomogeneity, the curvature of the background FLRW universe and a coupling term $-\alpha \dot{H}$. 

To clarify these curvature components, let us first imagine an observer at rest outside a Schwarzschild source of mass $M$ at radial coordinate $r_{\rm Sch}$ in the standard Schwarzschild coordinate system. The observer employs its natural adapted tetrad, where the spatial frame axes are along the spherical coordinate directions. The observed spacetime curvature is the projection of the curvature tensor upon the adapted tetrad of the observer and is given by 
\begin{equation}\label{N13a}
\mathcal{E} = {\rm diag}(-2\beta_{\rm Sch} , \beta_{\rm Sch}, \beta_{\rm Sch})\,, \qquad \mathcal{B} = 0\,,\qquad \mathcal{S} =  - \mathcal{E}\,, \qquad  \beta_{\rm Sch} = \frac{M}{r^3_{\rm Sch}}\,.
\end{equation}
In this connection, we note that Eq.~\eqref{N13} can be written as $\beta = M/\mathfrak{r}^3$, where $\mathfrak{r}$ is the Scwarzschild-like radial coordinate introduced in Eq.~\eqref{I10}. On the other hand, if $M = 0$, we find $\mathcal{E}$ is $qH^2$ times the  $3\times 3$ unit matrix, where $q H^2 = -\ddot{a}/a = -\dot{H} - H^2$, while $\mathcal{S}_1= \mathcal{S}_2 = \mathcal{S}_3 = H^2$.  The measured curvature is the sum of these independent components plus the interesting coupling term 
$-\alpha \,\dot{H} = \alpha\,(1+q) H^2$ that appears only in the diagonal gravitoelectric components of the measured curvature. The coupling term vanishes when the background is the de Sitter spacetime. 

Let us assume that the reference observer at radial coordinate $\bar{\rho}$ is far from the singularity such that  
\begin{equation}\label{N14}
\epsilon = \frac{M}{\bar{\rho} a(t)}\,, \qquad 0 < \epsilon \ll 1\,;
\end{equation}
then, we can write
\begin{equation}\label{N15}
\alpha = \epsilon +\frac{1}{2} \epsilon^2 + \cdots\,, \quad \beta= \frac{1}{\bar{\rho}^2 a^2(t)}(\epsilon - 3 \epsilon^2 + \cdots)\,, \quad \gamma = \frac{1}{\bar{\rho} a(t)}(\epsilon -  \epsilon^2 + \cdots)\,.
\end{equation}
 Henceforth, we keep only terms linear in the mass of the inhomogeneity.  For $\bar{\rho} \to \infty$, the fiducial observer is far from the inhomogeneity, which therefore has negligible influence and the radius of curvature of spacetime is in effect the Hubble radius $L_{\rm H} := c/H$. Otherwise, there is an interplay between $L_{\rm H}$ and the radius of curvature of the inhomogeneity $\ell$,  
\begin{equation}\label{N15a}
\beta \approx a^{-3}(t) \left(\frac{GM}{c^2 \bar{\rho}}\right) \frac{1}{\bar{\rho}^2} := \frac{1}{\ell^2}\,, \qquad  \ell = a(t)^{3/2}\,\left(\frac{GM}{c^2 \bar{\rho}}\right)^{-1/2} \, \bar{\rho}\,. 
\end{equation} 
The interplay between $\ell$ and $L_{\rm H}$ has a direct influence on the equations of motion discussed in detail in the next section. 
 
Finally, it is interesting to express the metric of the flat McVittie universe in terms of local Fermi coordinates. Using $\mathcal{E}_2 = \mathcal{E}_3$ and $\mathcal{S}_2 = \mathcal{S}_3$, we find
\begin{equation}\label{N16}
g_{\hat 0 \hat 0} = - (1+\gamma X^{\hat 1})^2  - \mathcal{E}_1\,(X^{\hat 1})^2 - \mathcal{E}_2\,(X^{\hat 2})^2 - \mathcal{E}_2\,(X^{\hat 3})^2 \,
\end{equation} 
  and 
\begin{equation}\label{N17}
g_{\hat i \hat j} = \delta_{\hat i \hat j} +\frac{1}{3} \,\left[
\begin{array} {ccc}
-\mathcal{S}_2 (X^{\hat 2})^2- \mathcal{S}_2 (X^{\hat 3})^2 & \mathcal{S}_2 X^{\hat 1}X^{\hat 2} & \mathcal{S}_2 X^{\hat 1}X^{\hat 3} \cr
\mathcal{S}_2 X^{\hat 1}X^{\hat 2} & -\mathcal{S}_1 (X^{\hat 3})^2 - \mathcal{S}_2 (X^{\hat 1})^2 & \mathcal{S}_1 X^{\hat 2}X^{\hat 3} \cr
\mathcal{S}_2 X^{\hat 1}X^{\hat 3} & \mathcal{S}_1 X^{\hat 2}X^{\hat 3} & -\mathcal{S}_1 (X^{\hat 2})^2- \mathcal{S}_2 (X^{\hat 1})^2 \cr
\end{array}
\right]\,,
\end{equation} 
while $g_{\hat 0 \hat i} = 0$. Taking advantage of the azimuthal symmetry about the direction toward the McVittie overdensity and introducing spherical polar coordinates $(\mathcal{R}, \Theta, \Phi)$, 
\begin{equation}\label{N18}
X^{\hat 1} = \mathcal{R} \cos\Theta\,, \qquad X^{\hat 2} = \mathcal{R} \sin\Theta \cos\Phi\,, \qquad X^{\hat 3} = \mathcal{R} \sin\Theta \sin\Phi\,,
\end{equation}
the Fermi metric becomes
\begin{align}\label{N19}
 ds^2 ={} & - [(1+\gamma \,\mathcal{R} \cos\Theta)^2 + \mathcal{R}^2(\mathcal{E}_1\,\cos^2 \Theta +\mathcal{E}_2\,\sin^2 \Theta)] \,dT^2 + d\mathcal{R}^2   \nonumber   \\ 
& +\mathcal{R}^2\,(1 - \tfrac{1}{3}\,\mathcal{R}^2\,\mathcal{S}_2)\,d\Theta^2 + \mathcal{R}^2 \sin^2\Theta\, [1 - \tfrac{1}{3}\,\mathcal{R}^2(\mathcal{S}_1\,\sin^2 \Theta +\mathcal{S}_2\,\cos^2 \Theta)]\,d\Phi^2\,.
\end{align} 

When $M = 0$, $\beta = \gamma = 0$,  $\mathcal{E}_1 = \mathcal{E}_2 = -\dot{H} - H^2 = q H^2$ and  $\mathcal{S}_1 = \mathcal{S}_2 = H^2$ for the flat FLRW model and the Fermi metric reduces to 
\begin{align}\label{N20}
 ds^2 ={} & - (1+ qH^2 \mathcal{R}^2)\,dT^2 + d\mathcal{R}^2 +\mathcal{R}^2\,(1 - \tfrac{1}{3}\,H^2\,\mathcal{R}^2)\,(d\Theta^2 + \sin^2\Theta\,d\Phi^2)\,.
\end{align} 
Indeed, very far from the overdensity, spacetime becomes homogeneous and isotropic and Eq.~\eqref{N19} approaches Eq.~\eqref{N20}. 
 
We now turn to a discussion of the timelike and null geodesic equations in the Fermi metric. A free test particle within the Fermi system has a unit 
 4-velocity vector $U^{\hat \mu}$,
\begin{equation}\label{G1}
U^{\hat \mu} := \frac{dX^{\hat \mu}}{ds} = \Gamma\,(1, \mathbf{V})\,,
\end{equation}
where $s$ is its proper time and  the Lorentz factor  $\Gamma$ is given by
\begin{equation}\label{G2}
\Gamma = \frac{1}{(- g_{\hat 0 \hat 0}-2\, g_{\hat 0 \hat i}\,V^{\hat i} - g_{\hat i \hat j}\,V^{\hat i}\,V^{\hat j})^{1/2}}\,. 
\end{equation}
The particle follows the timelike geodesic equation 
\begin{equation}\label{G3}
\frac{d^2 X^{\hat \mu}}{ds^2}+\Gamma^{\hat \mu}{}_{\hat \alpha \hat \beta}\,\frac{dX^{\hat \alpha}}{ds}\,\frac{dX^{\hat \beta}}{ds}=0\,.
\end{equation}
Separating this equation into its temporal and spatial components, we obtain the reduced geodesic equation~\cite{Chicone:2002kb}
\begin{equation}\label{G4}
\frac{d^2 X^{\hat i}}{dT^2}+\left(\Gamma^{\hat i}{}_{\hat \alpha \hat \beta}-\Gamma^{\hat 0}{}_{\hat \alpha \hat \beta}V^{\hat i} \right) \frac{dX^{\hat \alpha}}{dT}\frac{dX^{\hat \beta}}{dT}=0\,.
\end{equation}
In the immediate neighborhood of the reference observer, the space is Euclidean and Fermi velocity $\mathbf{V}$ of the test particle must satisfy the condition that $|\mathbf{V}| \le 1$ at $\mathbf{X}=0$. 

The reduced geodesic equation is valid for a null ray as well, provided
\begin{equation}\label{G5}
 g_{\hat 0 \hat 0} +2\, g_{\hat 0 \hat i}\,V^{\hat i} + g_{\hat i \hat j}\,V^{\hat i}\,V^{\hat j} = 0\,. 
\end{equation}

The Fermi metric has been expressed to second order in the spatial distance in our approximation scheme, cf. Eqs.~\eqref{N5}--\eqref{N7}; therefore, the connection coefficients in the Fermi frame are valid to linear order in the spatial distance. Indeed, we find
in the present case the nonzero components of the connection can be obtained from 
\begin{equation}\label{G6}
\Gamma^{\hat 0}_{\hat 0 \hat 0} = \frac{d\gamma}{dT} X^{\hat 1}\,, \quad \Gamma^{\hat i}_{\hat j \hat k} = -\frac{1}{3} (R_{\hat i \hat j \hat k \hat l} + R_{\hat i \hat k \hat j \hat l})\,X^{\hat l}\,, 
\end{equation}
\begin{equation}\label{G7}
\Gamma^{\hat 0}_{\hat 0 \hat 1} =  \gamma + (\mathcal{E}_1 + \gamma^2) X^{\hat 1}\,, \quad \Gamma^{\hat 0}_{\hat 0 \hat 2} = \mathcal{E}_2 X^{\hat 2}\,, \quad \Gamma^{\hat 0}_{\hat 0 \hat 3} = \mathcal{E}_3 X^{\hat 3}\,,
\end{equation}
\begin{equation}\label{G8}
\Gamma^{\hat 1}_{\hat 0 \hat 0} =  \gamma + (\mathcal{E}_1 + \gamma^2) X^{\hat 1}\,, \quad \Gamma^{\hat 2}_{\hat 0 \hat 0} = \mathcal{E}_2 X^{\hat 2}\,, \quad \Gamma^{\hat 3}_{\hat 0 \hat 0} = \mathcal{E}_3 X^{\hat 3}\,,
\end{equation}
using the symmetry of the Christoffel symbols.

\section{Peculiar Motions in Fermi Coordinates}

Using the Fermi-Walker transported frame $\bar{e}^{\mu}{}_{\hat \alpha}(\tau)$ along the world line of the reference observer, an approximate Fermi normal coordinate system has been established in its neighborhood. This system makes it possible to provide an invariant description of the motion of the test particles relative to the fiducial observer that occupies the origin of the spatial Fermi coordinates and its tetrad frame locally represents the rest frame of the gravitational inhomogeneity. More specifically, we study the local deviation of the world line of free test particles relative to the world line of the fiducial comoving observer. The resulting deviation is projected onto the Fermi-Walker transported tetrad frame of the reference observer. This approach furnishes an invariant characterization of local peculiar motions in cosmology. 

To simplify matters, we drop hats on the spatial position $\mathbf{X} = (X^1, X^2, X^3)$ and velocity $\mathbf{V} = (V^1, V^2 , V^3)$ of the free particle in the Fermi system. To express the reduced equation of geodesic motion (for a particle or a null ray), it is convenient to define
\begin{equation}\label{L1}
\mathcal{W}  := \mathcal{E}_1 X^1 V^1 + \mathcal{E}_2 X^2 V^2+ \mathcal{E}_3 X^3 V^3\,.
\end{equation}
Moreover, let us define the specific orbital angular momentum vector $\mathbf{L}$, 
\begin{equation}\label{L2}
L_i =  \epsilon_{ijk} X^j \, V^k\, 
\end{equation}
and
\begin{equation}\label{L3}
\mathbb{L}^1 := \mathcal{S}_1\, L_1\,, \qquad  \mathbb{L}^2 = \mathcal{S}_2\, L_2\,, \qquad  \mathbb{L}^3 = \mathcal{S}_3\, L_3\,.
\end{equation}
Then, using these quantities we define
\begin{equation}\label{L4}
\mathcal{V}_i :=  \epsilon_{ijk} \mathbb{L}^j \, V^k\,. 
\end{equation}

With these preliminaries, the equations of motion turn out to be
\begin{equation}\label{L5}
\frac{dV^1}{dT} + \gamma + (\mathcal{E}_1 + \gamma^2) X^1 -\frac{d\gamma}{dT}X^1V^1 - 2 \gamma (V^1)^2(1 + \gamma X^1) - 2 V^1 \mathcal{W} -\frac{2}{3}\mathcal{V}_1 = 0\,, 
\end{equation}
\begin{equation}\label{L6}
\frac{dV^2}{dT} + \mathcal{E}_2  X^2 -\frac{d\gamma}{dT}X^1V^2 - 2 \gamma (1 + \gamma X^1)V^1V^2   - 2 V^2 \mathcal{W} -\frac{2}{3}\mathcal{V}_2 = 0\,, 
\end{equation}
\begin{equation}\label{L7}
\frac{dV^3}{dT} + \mathcal{E}_3  X^3 -\frac{d\gamma}{dT}X^1V^3 - 2 \gamma (1 + \gamma X^1)V^1V^3  - 2 V^3 \mathcal{W} -\frac{2}{3}\mathcal{V}_3 = 0\,. 
\end{equation}
For $\gamma = 0$, the explicit forms of these equations are given in Appendix B. We must keep in mind that these equations are only approximately valid, since we have neglected higher-order curvature terms in the construction of the Fermi coordinate system~\cite{Chicone:2005vn}. 

\subsection{Homogeneous Case ($M = 0$)}

In this case, we are in the flat FLRW universe with $T = t$ and matrices $\mathcal{E}$ and $\mathcal{S}$ are proportional to the identity matrix with proportionality factors $qH^2$ and $H^2$, respectively. The equations of motion reduce essentially to the iteration of just one equation due to isotropy; that is, we have 
\begin{equation}\label{L8}
\frac{dV^1}{dt} + qH^2  X^1  - 2 V^1 \mathcal{W} -\frac{2}{3}\mathcal{V}_1 = 0\,, 
\end{equation}
\begin{equation}\label{L9}
\frac{dV^2}{dt} + qH^2  X^2 - 2 V^2 \mathcal{W} -\frac{2}{3}\mathcal{V}_2 = 0\,, 
\end{equation}
\begin{equation}\label{L10}
\frac{dV^3}{dt} + qH^2 X^3   - 2 V^3 \mathcal{W} -\frac{2}{3}\mathcal{V}_3 = 0\,. 
\end{equation}
Neglecting velocity terms in these equations, we find a ``Newtonian" equation of motion of the form
\begin{equation}\label{L11}
\frac{d^2 \mathbf{X}}{dt^2} + qH^2 \mathbf{X} = 0\,, 
\end{equation} 
which implies the existence of a relative cosmic tidal acceleration given by $-qH^2 \mathbf{X}$. The range of applicability of the equations of motion in this case is determined by the circumstance that Fermi coordinates are admissible for $|\mathbf{X}| \ll L_{\rm H}$, where $L_{\rm H} =c/H$ is the Hubble radius. 

To simplify matters, let us now consider motion in just one direction. In this case, we find from Eqs.~\eqref{L8}--\eqref{L10} that for each spatial direction, we have an equation of the form
\begin{equation}\label{L12}
\frac{d^2 \psi}{dt^2} + qH^2 (1- 2\,\dot{\psi}^2) \psi = 0\,, 
\end{equation} 
where $\dot{\psi} := d\psi/dt$. This equation has an exact solution involving uniform rectilinear motion at the constant critical speed $V_{\rm c} = c/\sqrt(2) \approx 0.707 c$; that is,  
\begin{equation}\label{L13}
\psi (t) = \psi (t_i) \pm\, (2)^{-1/2} (t-t_i)\,, 
\end{equation}
where $t_i$ is the initial instant of time. Moreover, Eq.~\eqref{L12} is invariant under $\psi \to -\psi$ and has a rest point at $(\psi, \dot{\psi}) = (0, 0)$, which means that there is no motion once the initial speed is zero at $\psi = 0$. 

Let us now imagine that $qH^2 := K$ is effectively a constant. In this case, Eq.~\eqref{L12} is  autonomous, completely integrable and its first integral is given by
\begin{equation}\label{L14}
\dot{\psi}^2 + (\tfrac{1}{2} - \dot{\psi}_i^2) e^{2K\psi^2} =  \frac{1}{2}\,, 
\end{equation}
where we have assumed that at $\psi = 0$, $\dot{\psi} = \dot{\psi}_i$ is the initial speed.   Let us note that in the current benchmark model, $q_0 \approx -0.55$~\cite{Dodelson:2003ft,Weinberg:2008zzc,Amendola:2015ksp,Baumann:2022mni}, so that $K_0 < 0$. 

Consider motion in the positive $\psi$ direction; that is, the free test particle moves radially away from the fiducial observer. For $q < 0$, the free particle accelerates (decelerates) if the initial speed is less (greater) than the critical speed; that is, the character of the motion is toward the exact solution, which acts as a local attractor~\cite{Chicone:2002kb}. 
On the other hand, for $q > 0$, the free particle accelerates (decelerates) if the initial speed is greater (less) than the critical speed; hence, the nature of the motion is away from the exact solution~\eqref{L13}. 

The nature of the motion can be further elucidated in both cases if we regard Eq.~\eqref{L14} as the constant total energy relation for a hypothetical one-dimensional motion in classical mechanics with effective potential energy $V_{eff}(\psi) =  (\tfrac{1}{2} - \dot{\psi}_i^2) \exp{(2K\psi^2)}$~\cite{Mashhoon:2020tha}.  Hence, motion is confined to the region where $2 \,V_{eff}(\psi) \le 1$.   In particular, if the initial speed is more than zero but less than the critical speed and $K > 0$, the motion is periodic and the particle moves back and forth between turning points given by  $V_{eff}(\psi) = 1/2$, as in Figure 1. However, the character of the motion changes drastically if the initial speed is above the critical speed; that is, the free particle accelerates away from the critical speed and toward the speed of light, though this situation would be moderated by the presence of higher-order curvature terms that we have neglected in our construction of the Fermi coordinate system~\cite{Chicone:2005vn}. 

Let us now return to the case where $K < 0$. In the de Sitter limit of FLRW cosmology, $q = -1$ and $H = (\Lambda/3)^{1/2} =  {\rm constant}$; hence, $K < 0$ is constant. In this case all local peculiar motions asymptotically approach the rectilinear motion with critical speed $\approx 0.7 c$. The expansion of the universe at the present cosmic epoch appears to be dominated by the cosmological constant; therefore, we expect peculiar motions to proceed eventually toward the critical speed. In this connection, it is not clear how to relate phenomena in Fermi coordinates centered on other galaxies to our measurements of the properties of large-scale peculiar motions of clusters of galaxies.

We now turn to the consequences of inhomogeneity due to the presence of mass $M$.

%%%%%%%%%%%%%
\begin{figure}\label{fig:1}
    \centering
    \includegraphics[scale=0.35]{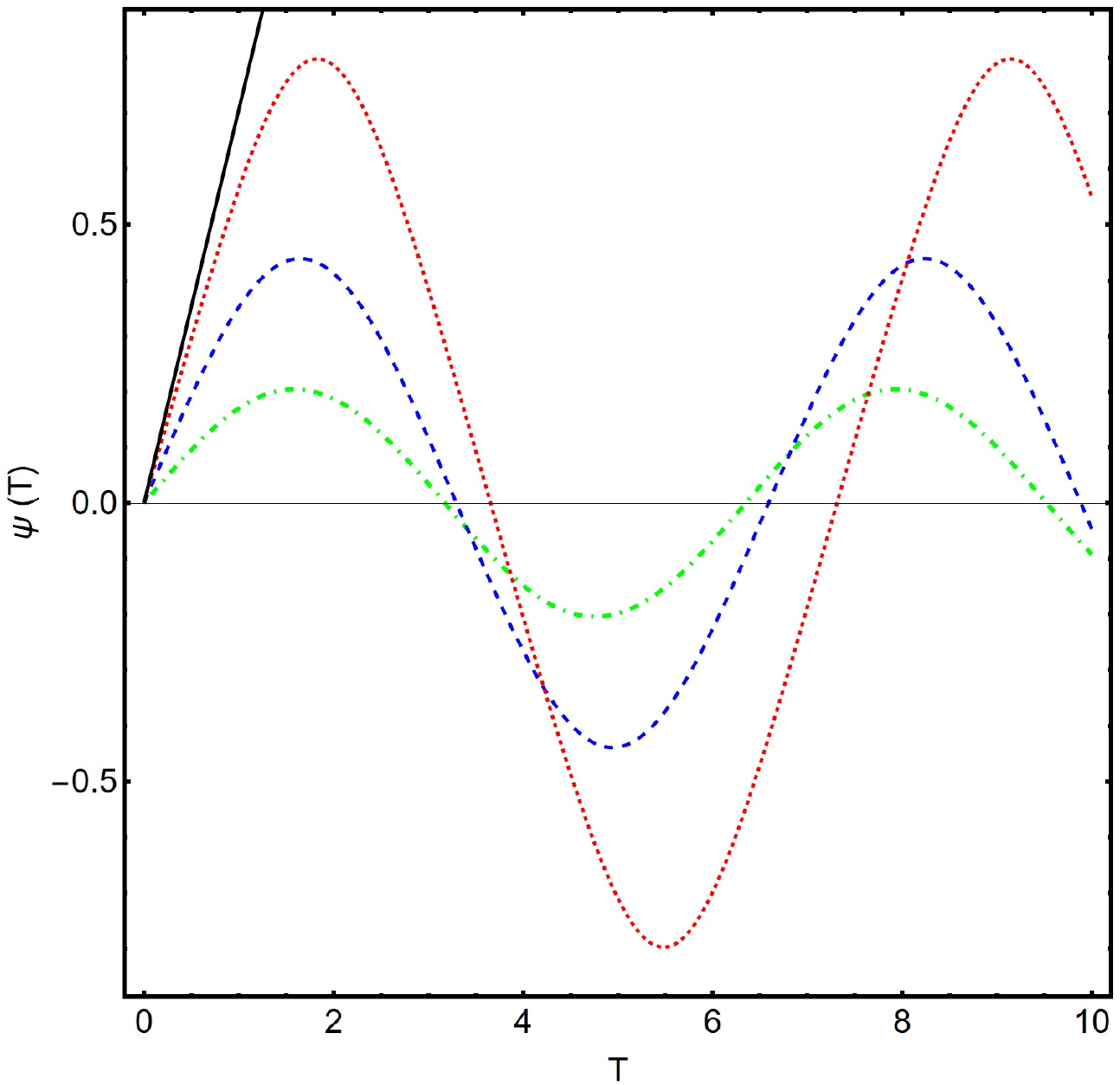}\caption{The spatial Fermi coordinate $\psi(T)$, which satisfies the autonomous ordinary differential equation~\eqref{L12} for constant $qH^2$, is plotted versus $T = ct + {\rm constant}$, where $t$ is cosmic time. Here, all lengths are expressed in units of a constant length $\lambda$. Equation~\eqref{L12} under consideration here is not explicitly dependent upon the temporal variable; therefore, we write its dependence upon cosmic time $t$ as $T = ct + {\rm constant}$ to indicate that our result in this figure for $T: 0 \to 10$ is independent of any specific cosmic epoch. We numerically solve Eq.~\eqref{L12} for $qH^2 = K = 1$, where $K$ is expressed in units of $\lambda^{-2}$. The boundary conditions are that at $T= 0$, $\psi = 0$ and the initial speed $\dot{\psi}_i$ takes different nonzero values. For  $\dot{\psi}_i$ equal to the critical speed ($\approx 0.7$), the exact solution is a straight line (solid black); however, for initial nonzero speeds below the critical speed, the solutions are all periodic, as illustrated here for
$\dot{\psi}_i = 0.6$ (red dot), $\dot{\psi}_i = 0.4$ (blue dash) and $\dot{\psi}_i = 0.2$ (green dot-dash).}
\end{figure}
%%%%%%%%%%%%%%

%%%%%%%%%
%%%%%%%%%%

\subsection{Inhomogeneous Case ($M \ne 0$)}

To simplify matters, let us first drop all terms that are second order in $M$. Then, Eqs.~\eqref{N2} and~\eqref{N15} imply 
\begin{equation}\label{M1}
\frac{d\gamma}{dT} = - \frac{2M}{\bar{\rho}^2 a^2} H\,. 
\end{equation} 
Similarly, we have
\begin{equation}\label{M2}
\alpha = \frac{M}{\bar{\rho} \,a(t)}\,, \qquad \beta = \frac{M}{\bar{\rho}^3 a^3(t)}\,, \qquad \gamma = \frac{M}{\bar{\rho}^2 a^2(t)}\,. 
\end{equation}
Moreover, we recall from Eq.~\eqref{N12a} that $\mathcal{E}_1 = -2\,\beta + qH^2 -\alpha \dot{H}$ and  $\mathcal{E}_2 = \mathcal{E}_3 = \beta+ qH^2 -\alpha \dot{H}$. Equation~\eqref{L5} for motion purely along the $X^1$ direction then implies
\begin{equation}\label{M3}
\frac{dV^1}{dT} + (- 2 \beta + qH^2 - \alpha \dot{H}) [ 1 - 2 (V^1)^2]  X^1  +\frac{M}{\bar{\rho}^2 a^2} [ 1 + 2 H X^1 V^1 - 2 (V^1)^2] = 0\,. 
\end{equation}
On the other hand, for motion purely in the $X^2$ or $X^3$ directions, we get 
\begin{equation}\label{M4}
\frac{dV^2}{dT} + (\beta + qH^2 - \alpha \dot{H}) [ 1 - 2 (V^2)^2]  X^2 = 0\, 
\end{equation}
and similarly for $X^3$. We note that Eq.~\eqref{M4} has the standard form~\eqref{L12}, except that $qH^2$ is replaced by $\beta + qH^2 - \alpha \dot{H}$, which is a sum of the curvature ($\beta$) due to $M$, the background flat FLRW curvature ($qH^2$)  and a coupling term proportional to $\dot{H}$. 

%%%%%%%%%%%
%%%%%%%%%%%
\begin{figure}\label{fig:2}
 \centering
    \includegraphics[scale=0.3]{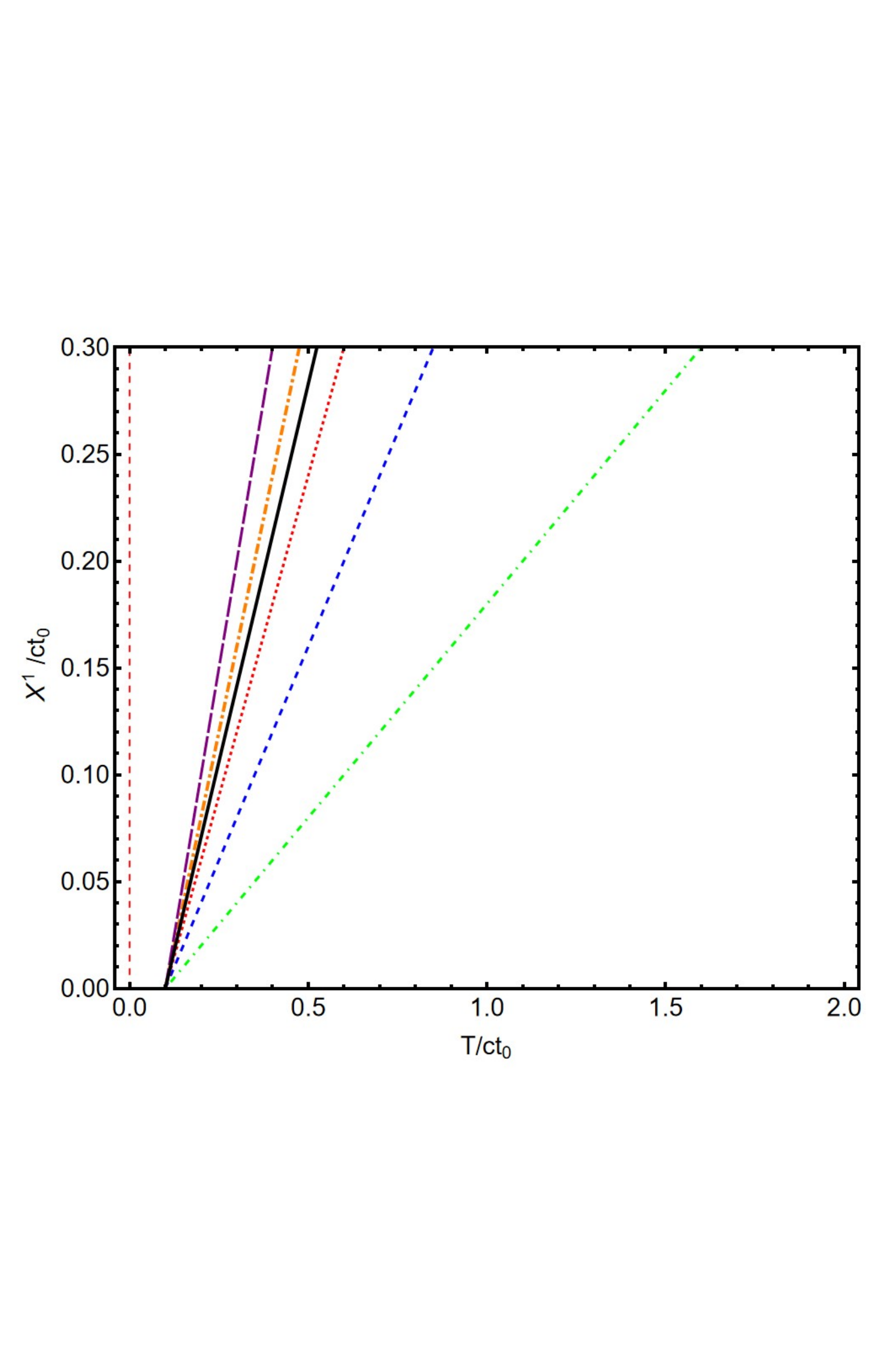}\caption{Plot of $\bar{X}^1$ versus $\bar{T}$ obtained from the numerical integration of Eq.~\eqref{M6} with parameters given in Eq.~\eqref{M10} that isolate the curvature coupling term. The integration extends over the range $\bar{T} = 0.1 \to 2$ with initial conditions $\bar{X}^1 = 0$ and initial speed $V^1$. We show the plots for $V^1= 0.2$ (green dot-dash), $V^1= 0.4$ (blue dash), $V^1= 0.6$ (red dot), $V^1= 1/\sqrt{2}$ (solid black), $V^1= 0.8$ (orange dot-dash) and the null geodesic $V^1= 1$ (purple em-dash).}
\end{figure}
%%%%%%%%%
%%%%%%%%
%%%%%%%%%%
%%%%%%%%%%
 \begin{figure}\label{fig:3}
  \centering
    \includegraphics[scale=0.3]{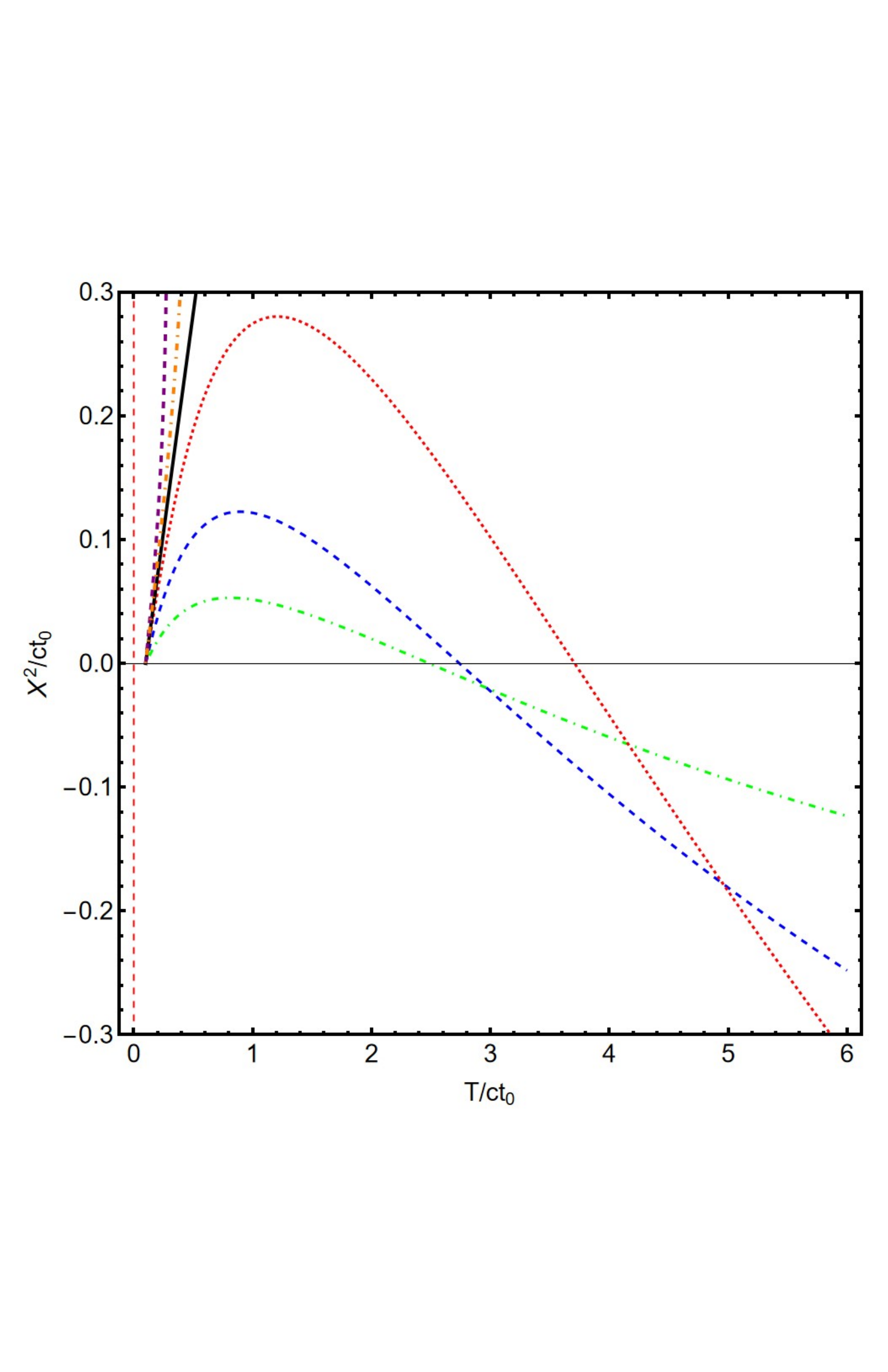}\caption{Plot of $\bar{X}^2$ versus $\bar{T}$ obtained from the numerical integration of Eq.~\eqref{M7} with parameters given in Eq.~\eqref{M11}. The integration extends over the range $\bar{T} = 0.1 \to 6$ with initial conditions $\bar{X}^2 = 0$ and initial speed $V^2$. We show the plots for $V^2= 0.2$ (green dot-dash), $V^2= 0.4$ (blue dash), $V^2= 0.6$ (red dot), $V^2= 1/\sqrt{2}$ (solid black line), $V^2= 0.8$ (orange dot-dash) and the null geodesic 
    $V^2= 1$ (purple dash).} 
\end{figure}
%%%%%%%%
%%%%%%%

If the fiducial observer is sufficiently close to the inhomogeneity, it is clear from the equations of motion that the time-dependent background FLRW cosmology has negligible influence on the motion. For example, let us choose a galactic mass 
$M = 10^{11} M_{\odot}$ embedded in the background flat model and let the comoving observer be at a radial distance of $\bar{\rho} = 10$ kpc. Then, at the present epoch,  $\ell \approx 5 \times 10^{25}$ cm and $\ell/L_{\rm H} \approx 5 \times 10^{-3}$. On the other hand, if the comoving observer is about a hundred times further away, $\ell$ and $L_{\rm H}$ are comparable and the background must be taken into consideration. That is, the equations of motion must be solved along with the evolution equations for the background model. To simplify matters, let us assume that the background is the matter-only Einstein-de Sitter universe with $a(t) = (t/t_0)^{2/3}$, where $t_0$ is the current age of the universe. Working to first order in $GM/(c^2\, \bar{\rho}) \ll1$, we can then integrate Eq.~\eqref{N2} and express the result as
\begin{equation}\label{M5}
\bar{T} := \frac{T}{ct_0}\,, \quad \bar{t} := \frac{t}{t_0}\,,\quad \bar{T} = \bar{t} - 3 \left(\frac{GM}{c^2\bar{\rho}}\right) \,\bar{t}^{\,1/3}\,, \quad  \bar{t} = \bar{T} + 3  \left(\frac{GM}{c^2\bar{\rho}}\right) \,\bar{T}^{1/3}\,.
\end{equation}
Next, let us define $\bar{X}^i := X^i/(ct_0)$, so that $\bar{V}^i := d \bar{X}^i/d \bar{T} = V^i$  and write Eqs.~\eqref{M3} and~\eqref{M4} in dimensionless form  as
\begin{equation}\label{M6}
\frac{d^2 \bar{X}^1}{d\bar{T}^2} + \mathbb{K}_1 [ 1 - 2 (V^1)^2]  \bar{X}^1  +\frac{GM}{c^2 \bar{\rho}}\,\frac{ct_0}{\bar{\rho}}\,\bar{T}^{-4/3} [ 1 + \tfrac{4}{3}\bar{T}^{-1} \,\bar{X}^1 V^1 - 2 (V^1)^2] = 0\,, 
\end{equation}
\begin{equation}\label{M7}
\frac{d^2 \bar{X}^2}{d\bar{T}^2} + \mathbb{K}_2 [ 1 - 2 (V^2)^2]  \bar{X}^2 = 0\,, 
\end{equation} 
where  
\begin{equation}\label{M8}
\mathbb{K}_1 = \frac{2}{9}\bar{T}^{-2}\left(1 - 9\, \frac{GM}{c^2 \bar{\rho}}\,\frac{c^2t_0^2}{\bar{\rho}^2} - 3\,\frac{GM}{c^2 \bar{\rho}}\,\bar{T}^{-2/3}\right)\,, 
\end{equation}   
\begin{equation}\label{M9}
\mathbb{K}_2 = \frac{2}{9}\bar{T}^{-2}\left(1 +\frac{9}{2}\, \frac{GM}{c^2 \bar{\rho}}\,\frac{c^2t_0^2}{\bar{\rho}^2} - 3\,\frac{GM}{c^2 \bar{\rho}}\,\bar{T}^{-2/3}\right)\,.
\end{equation}
We numerically integrate Eq.~\eqref{M6} with initial conditions that $\bar{X}^1 = 0$ at $\bar{T} = 0.1$ with various initial speeds $V^1 = 0.2, 0.4,  0.6, 1/\sqrt{2}, 0.8, 1$ and  parameters
\begin{equation}\label{M10}
M \approx 3 \times 10^{11} M_{\odot}\,, \qquad \bar{\rho} \approx 1500~ {\rm kpc}\,, \qquad  \frac{GM}{c^2 \bar{\rho}} = 10^{-8}\,, \qquad \frac{ct_0}{\bar{\rho}} = \frac{1}{3} \times 10^4\,.
\end{equation}
The result is illustrated in Figure 2. 

Next, we numerically integrate Eq.~\eqref{M7} with initial conditions $\bar{X}^2 = 0$ at $\bar{T} = 0.1$ with various initial speeds $V^2 = 0.2, 0.4,  0.6, 1/\sqrt{2}, 0.8, 1$ and  parameters
\begin{equation}\label{M11}
M \approx 10^{11} M_{\odot}\,, \qquad \bar{\rho} \approx 500~ {\rm kpc}\,, \qquad  \frac{GM}{c^2 \bar{\rho}} =10^{-8}\,, \qquad \frac{ct_0}{\bar{\rho}} = 10^4\,.
\end{equation}
Figure 3 contains our numerical results.

It is interesting to compare and contrast our results for the flat McVittie model with a somewhat similar LTB model. 

%%%%%%%%%
%%%%%%%%%%
%%%%%%%%%

\section{Peculiar Motions in a Simple LTB Model}

To introduce the new model, let us recall that the metric of a general spherically symmetric spacetime in comoving coordinates $x^\mu = (t, r, \theta, \phi)$ can be written in the form 
\begin{equation}\label{P1}
ds^2 = - A^2(t, r)\, dt^2 + B^2(t, r)\, dr^2 + R^2(t, r)\, d\Omega^2\,, 
\end{equation}
as in Eq.~\eqref{I19}. In the LTB model, the metric functions are given by
\begin{equation}\label{P2}
A(t, r) = 1, \qquad B(t, r) = [1+ 2\,\mathbb{E}(r)]^{-1/2} \,\frac{\partial R}{\partial r}\,,
\end{equation}
where $\mathbb{E}(r) > -\tfrac{1}{2}$. The energy-momentum tensor for dust can be expressed as
\begin{equation}\label{P3}
T_{\mu \nu} = \mu_{\rm LTB}(t, r)\, u_\mu u_\nu\,, \qquad u^\mu = \delta^\mu_0\,.
\end{equation}

The gravitational field equations in this case imply
\begin{equation}\label{P4}
\mathbb{E}(r) = \frac{1}{2} \left(\frac{\partial R}{\partial t}\right)^2 -\frac{G\mathcal{M}(r)}{R} -\frac{1}{6} \Lambda R^2\,
\end{equation}
and
\begin{equation}\label{P5}
\frac{d\mathcal{M}(r)}{dr} = 4 \pi \,\mu_{\rm LTB}(t, r)\,R^2 \,\frac{\partial R}{\partial r}\,,
\end{equation}
where $\Lambda$ is the cosmological constant, $\mathbb{E}(r)$ is the energy per unit mass of a spherical shell of dust of radius $r$ and $\mathcal{M}(r)$ is the mass-energy within a sphere of radius $r$. 
In spherically symmetric spacetimes, the invariant Misner-Sharp mass~\cite{Misner:1964je, HeMi, CaMc}  $m(t, r)$ defines the amount of mass-energy within a radius $r$ at time $t$ and is given explicitly by~\cite{GlaMa, Mashhoon:1979tt}
\begin{equation}\label{P6}
m(t, r) = \frac{1}{2} R \left[ 1 + \left(\frac{1}{A}\frac{\partial R}{\partial t}\right)^2 -  \left(\frac{1}{B}\frac{\partial R}{\partial r}\right)^2\right]\,.
\end{equation}
It follows that the Misner-Sharp mass in this case is independent of time and is given by
\begin{equation}\label{P7}
m_{\rm LTB} = \mathcal{M}(r)\,.
\end{equation}

The LTB model generalizes the standard Friedmann model (i.e., zero-pressure FLRW universe). The metric of the FLRW universe can be expressed as 
\begin{equation}\label{P8}
ds^2|_{\rm FLRW} = -dt^2 + a^2(t) \left[\frac{dr^2}{1-\tfrac{k}{R_0^2}\,r^2} + r^2 d\Omega^2\right],
\end{equation}
where $k = 1$, $-1$, or $0$, for the closed, open, or flat model, respectively, and $R_0$ is a constant length scale. The LTB metric reduces to Eq.~\eqref{P8} provided we assume
\begin{equation}\label{P9}
R(t, r) = a(t)\, r\,, \qquad 2\,\mathbb{E} = -\frac{k}{R_0^2}\,r^2\,, \qquad \mathcal{M}(r) = \mathcal{M}_0\, r^3\,;
\end{equation}
then, $\mu_{\rm LTB}(t, r) \to \mu_{\rm F}(t)$, where $\mu_{\rm F}(t)$ is the energy density of the Friedmann universe, and $\mathcal{M}_0$ is a constant given by
\begin{equation}\label{P10}
 \mathcal{M}_0 = \frac{4 \pi}{3}\,a^3(t)\, \mu_{\rm F}\,.
\end{equation}
The pressure vanishes in the Friedmann model; therefore, Eq.~\eqref{5} implies that $\mu_{\rm F} a^3(t)$ is constant, in agreement with Eq.~\eqref{P10}. 

For our specific LTB model, we assume $\mathbb{E} = \Lambda = 0$. In this case, the general solution of Eq.~\eqref{P4} is given by
\begin{equation}\label{P11}
R(t, r) = [\tfrac{9}{2}G\mathcal{M}(r)]^{1/3}\, [t - t_B(r)]^{2/3}\,,
\end{equation}
where $t_B(r)$ denotes the time that the big bang singularity takes place for a given radial coordinate $r$. It remains to specify $\mathcal{M}$ and $t_B(r)$. We assume
\begin{equation}\label{P12}
\mathcal{M} = \mathbb{M} \left(1 + \frac{r^3}{r_0^3}\right)\,,\qquad \mathbb{M} = \mathcal{M}_0\, r_0^3\,, \qquad  t_B(r) = \frac{\vartheta}{1 + \frac{r^3}{r_0^3}}\,,
\end{equation}
where $\mathbb{M}$, $r_0$ and $\vartheta$ are constants. We find
\begin{equation}\label{P13}
 \frac{\partial R}{\partial r}  = (\tfrac{9}{2}G\mathcal{M}_0)^{1/3}\,\left(1 + \frac{r_0^3}{r^3}\right)^{-2/3} \frac{t + t_B(r)} {[t - t_B(r)]^{1/3}}\,.
\end{equation}
It then follows from Eq.~\eqref{P5} that
\begin{equation}\label{P14}
 6 \pi G\, \mu_{\rm LTB}  = \frac{1}{t^2 - t_B^2}\,.
\end{equation}
 
The LTB spacetime is singular at $t = t_B = \vartheta /(1 + r^3/r_0^3)$, where $ \mu_{\rm LTB}$ diverges. At the center of spherical symmetry $r = 0$, the singularity occurs at time $t = \vartheta$, while at $r = \infty$, the singularity occurs at $t = 0$. For $r >> r_0$, $\mathcal{M} \to \mathcal{M}_0r^3$, $t_B(r) \propto (r/r_0)^{-3}$ and the spacetime asymptotically approaches the flat Friedmann model known as the Einstein-de Sitter universe where $t_B \to 0$, as expected. 

The LTB model thus represents a spherically symmetric inhomogeneity embedded in a Friedmann background, just as the McVittie model represents a Schwarzschild solution embedded in a FLRW background. 

In our simple LTB model, we note that
\begin{equation}\label{P15}
(\mu_{\rm LTB})_{; \alpha}{}^{; \alpha}= \frac{1}{\sqrt{-g}} \frac{\partial}{\partial x^\alpha}\left(\sqrt{-g} g^{\alpha \beta}\frac{\partial \mu_{\rm LTB}}{\partial x^\beta}\right)\,, \qquad \lim_{r \to 0} (\mu_{\rm LTB})_{; \alpha}{}^{; \alpha} \to \infty\,. 
\end{equation}
The source of this divergence is simply due to the introduction of the point mass $\mathbb{M}$ at $r=0$ in our simple model in analogy with the central Schwarzschild mass $M$ in the McVittie case; that is, a discontinuity is thereby created in the gradient of mass density at $r= 0$, which leads to the divergence. That this is not a spacetime curvature singularity has been elucidated in~\cite{Krasinski:2010pfm}.  

To study local peculiar motions in this LTB spacetime and compare our results with the flat McVittie model, we make use of the quasi-inertial Fermi normal coordinate system. 

%%%%%%%%%%%
%%%%%%%%%%%
%%%%%%%%%%%

\subsection{Fermi Coordinate System}

We now proceed to construct a Fermi normal coordinate system centered on a fiducial preferred observer in our simple LTB model with metric 
\begin{equation}\label{w1}
 ds^2 = -dt^2 + R'^2(t, r) dr^2 + R^2(t, r)(d\theta^2 + \sin^2 \theta \,d\phi^2)\,.
\end{equation}
Henceforth, a prime indicates partial derivative with respect to $r$, so that $R' = \partial{R} / \partial{r}$ given in Eq.~\eqref{P13}, and $\dot{R} = \partial{R} / \partial{t}$, as before.  The observer is located at $(\bar{r}, \bar{\theta}, \bar{\phi})$ and has an adapted orthonormal tetrad frame $\chi^{\mu}{}_{\hat \alpha}$,
\begin{equation}\label{w2}
\chi^{\mu}{}_{\hat 0} = \delta^\mu_0\,,  \quad \chi^{\mu}{}_{\hat 1} := \frac{1}{\bar{R}'}\, \delta^\mu_{1}\,,\quad \chi^{\mu}{}_{\hat 2} := \frac{1}{\bar{R}}\, \delta^\mu_{2}\,, \quad \chi^{\mu}{}_{\hat 3} := \frac{1}{\bar{R}\,\sin \bar{\theta}}\, \delta^\mu_{3}\,.
\end{equation} 
An arbitrary comoving observer follows a geodesic and the adapted tetrad frame field is parallel transported along the world line of the observer; hence,  
\begin{equation}\label{w3}
\frac{D \chi^{\mu}{}_{\hat \alpha}}{d\tau} = \tilde{\Phi}_{\hat \alpha}{}^{\hat \beta} \, \chi^{\mu}{}_{\hat \beta}\,
\end{equation}
such that $\tilde{\Phi}_{\hat \alpha \hat \beta} = 0$. Next, we calculate
\begin{equation}\label{w4}
\tilde{R}_{\hat \alpha \hat \beta \hat \gamma \hat \delta} = R_{\mu \nu \rho \sigma}\,\chi^{\mu}{}_{\hat \alpha}\,\chi^{\nu}{}_{\hat \beta}\,\chi^{\rho}{}_{\hat \gamma}\,\chi^{\sigma}{}_{\hat \delta}\,
\end{equation}
and find the electric, magnetic and spatial components of curvature of our specific LTB model.  

In the LTB universe, the measured curvature has the same structure as in the McVittie universe; that is, 
\begin{equation}\label{w5}
\tilde{\mathcal{E}} = {\rm diag}(\tilde{\mathcal{E}}_1, \tilde{\mathcal{E}}_2, \tilde{\mathcal{E}}_3)\,, \qquad \tilde{\mathcal{B}} = 0\,,\qquad \tilde{\mathcal{S}} = {\rm diag}(\tilde{\mathcal{S}}_1, \tilde{\mathcal{S}}_2, \tilde{\mathcal{S}}_3)\,,
\end{equation}
where $\tilde{\mathcal{E}}$, $\tilde{\mathcal{B}}$ and $\tilde{\mathcal{S}}$ denote, as before,  the gravitoelectric,  gravitomagnetic and spatial components of the Riemann curvature tensor  as measured by the reference observer, respectively. Specifically, we find
\begin{align}
    &\tilde{\mathcal{E}}_{1}=-\frac{\ddot{\bar{R}}'}{\bar{R}'}= \frac{2}{9}\,\frac{t-5\bar{t}_B}{(t-\bar{t}_B)^2(t+\bar{t}_B)}\,,\label{w6}
    \\
    &\tilde{\mathcal{E}}_{2}= \tilde{\mathcal{E}}_{3}=-\frac{\ddot{\bar{R}}}{\bar{R}}=\frac{2}{9}\,\frac{1}{(t-\bar{t}_B)^2}\,,\label{w7}
    \\
    &\tilde{\mathcal{S}}_{1}=\frac{\dot{\bar{R}}^2}{\bar{R}^2}= \frac{4}{9}\,\frac{1}{(t-\bar{t}_B)^2}\,,  \label{w8}
    \\
    &\tilde{\mathcal{S}}_{2}= \tilde{\mathcal{S}}_{3}=\frac{\dot{\bar{R}} \dot{\bar{R}}'}{\bar{R} \bar{R}'}= \frac{4}{9}\,\frac{t-2\bar{t}_B}{(t-\bar{t}_B)^2(t+\bar{t}_B)}\,.  \label{w9}
\end{align}
These results clearly exhibit the curvature singularity of the LTB model at $t = \bar{t}_B = \vartheta /(1 + \bar{r}^3/r_0^3)$. 

It is remarkable that  the measured curvature components are independent of mass $\mathbb{M}$. This circumstance appears to be in contrast to Eqs.~\eqref{N11}--\eqref{N13} that hold in the McVittie spacetime. While the inhomogeneity in the McVittie spacetime is due to the embedding of mass $M$ in the FLRW background, the radial inhomogeneity in the LTB spacetime exists regardless of the presence of $\mathbb{M}$.

In the limit that $\bar{r}/r_0$ approaches infinity, $\bar{t}_B \to 0$ and we find
\begin{equation}\label{w10}
\tilde{\mathcal{E}}_1= \tilde{\mathcal{E}}_2= \tilde{\mathcal{E}}_3 = \frac{2}{9\,t^2}\,, \qquad \tilde{\mathcal{S}}_1= \tilde{\mathcal{S}}_2= \tilde{\mathcal{S}}_3 = \frac{4}{9\,t^2}\,,
\end{equation}
that correspond to $qH^2$ and $H^2$, respectively, characteristics of the Einstein-de Sitter model with $H = 2/(3\,t)$ and $q = 1/2$. 

The Fermi metric for the LTB model can be simply obtained from Eq.~\eqref{N19} by letting $\gamma = 0$, $T = t$ and  $(\mathcal{E} \to \tilde{\mathcal{E}}$, $\mathcal{S} \to \tilde{\mathcal{S}})$.

\subsection{Geodesic Motion}

Comoving observers in the LTB spacetime follow geodesics; therefore, the equations of motion in this case are the same as those given in Appendix B, except that we must let $T = t$,  $\mathcal{E} \to \tilde{\mathcal{E}}$ and $\mathcal{S} \to \tilde{\mathcal{S}}$. To simplify matters, we only consider motion in a single direction. Inspection of the equations of motion reveals that for motion purely in the $X^1$ direction, we get 
\begin{equation}\label{w11}
 \frac{d^2X^1}{dt^2} +  \tilde{\mathcal{E}}_1\,[1-2(V^1)^2]\,X^1 = 0\,,
\end{equation}
and for motion purely in the $X^2$ direction
\begin{equation}\label{w12}
 \frac{d^2X^2}{dt^2} +  \tilde{\mathcal{E}}_2\,[1-2(V^2)^2]\,X^2 = 0\,,
\end{equation}
while the corresponding equation for $X^3$ will be the same as the one for $X^2$. We note that $\tilde{\mathcal{E}}_1$ given by Eq.~\eqref{w6} starts out at $-\infty$ for $t = \bar{t}_B$, rises rapidly above zero, has a maximum and then falls off to zero as $t \to \infty$, while  $\tilde{\mathcal{E}}_2$ given by Eq.~\eqref{w7} starts out at $+\infty$ for $t = \bar{t}_B$, decreases rapidly and goes to zero as $t \to \infty$.  Uniform motion with speed $c/\sqrt{2}$ is an exact solution in both equations, as noted before. 
We have solved Eqs.~\eqref{w11} and~\eqref{w12} numerically. The results for Eq.~\eqref{w11} are presented in Figure 4. For initial speeds below and above the critical speed, the motion turns away from the straight line at the critical speed. We find essentially the same outcome for Eq.~\eqref{w12}. Despite the differences between $\tilde{\mathcal{E}}_1$ and $\tilde{\mathcal{E}}_2$, the numerical results are qualitatively the same.

%%%%%%%%%%%%%%%
\begin{figure}\label{fig:4}
 \centering
    \includegraphics[scale=0.35]{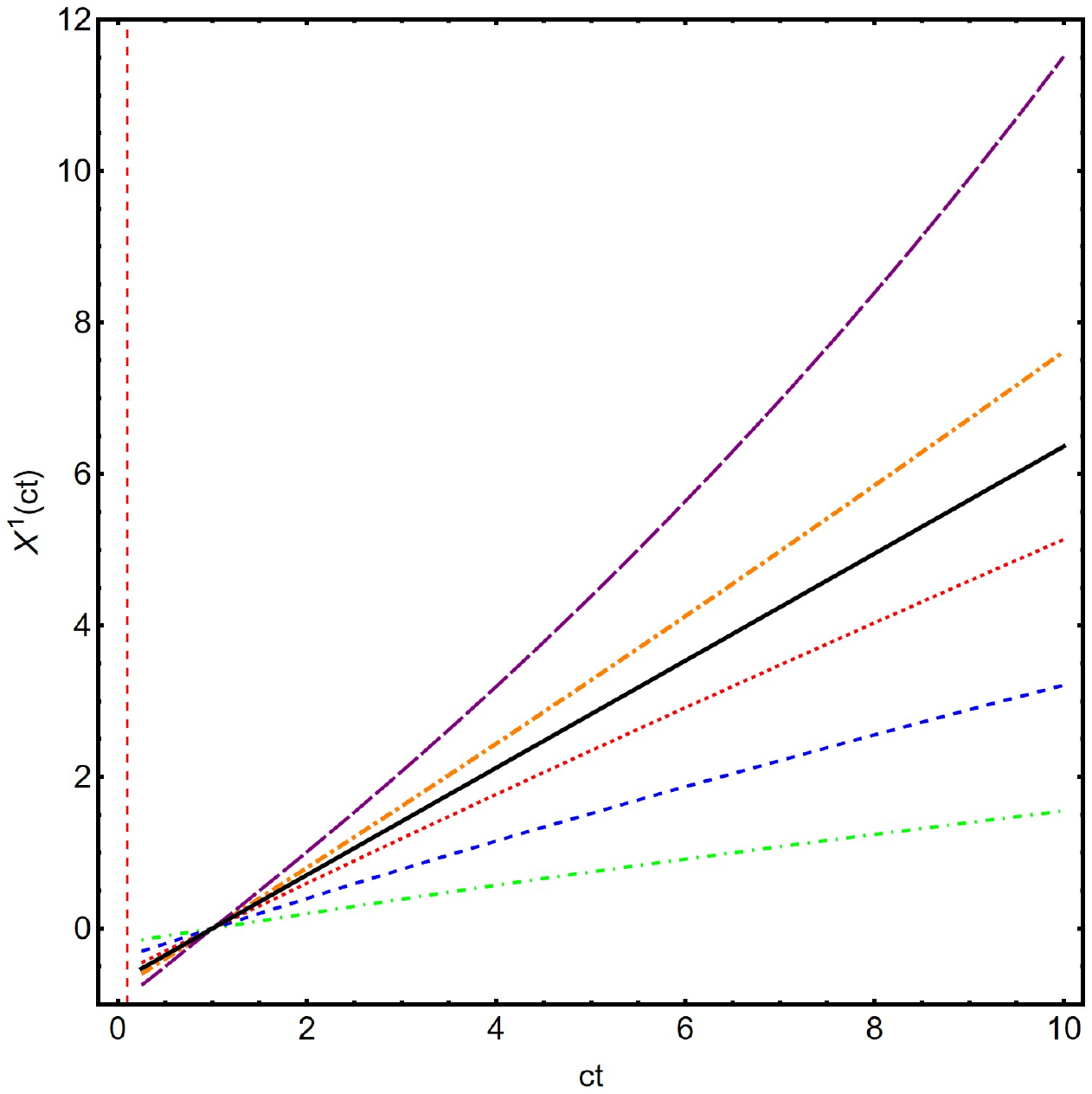}\caption{Plot of $X^1$ given by Eq.~\eqref{w11}  versus $ct$ from $ct = 1$  to $ct = 10$ with $c\bar{t}_{B}=0.1$. Here, all lengths are expressed in units of a constant length $\lambda$. We represent $c\bar{t}_{B}=0.1$ by a vertical red dotted line.The exact solution at the critical speed ($\approx 0.7$) is a straight line (solid black). For initial nonzero speeds below the critical speed, the solutions turn downward, as illustrated here for 
$V^{1}(ct =1)=0.6$ (red dot), $V^{1}(ct =1)=0.4$ (blue dash) and $V^{1}(ct =1)=0.2$ (green dot-dash). For initial speeds above the critical speed, the solutions turn upward, as illustrated here for 
$V^{1}(ct =1)=0.8$ (orange dot-dash) and the null geodesic $V^{1}(ct =1)= 1$ (purple em-dash). 
}
   \end{figure}
%%%%%%%%%
%%%%%%%%%
%%%%%%%%

\section{DISCUSSION}

The invariantly defined quasi-inertial Fermi normal coordinate system is ideally suited to the description of measurements of a local observer that carries an orthonormal spatial frame consisting of ideal nonrotating (i.e., Fermi-Walker transported) test gyroscopes. In particular, this system can be used to determine the influence of the expansion of the universe on local physics. In the FLRW cosmology, spatial distances within the Fermi system must be very small in comparison with the Hubble radius. Recently, conformal Fermi coordinates have been introduced as a generalization of Fermi coordinates that are valid on suprerhorizon scales within the context  of FLRW cosmology~\cite{Pajer:2013ana, Dai:2015rda}. 

Employing Fermi normal coordinate systems, we have studied tidal dynamics and equations of motions of free test particles relative to reference comoving observers in the flat McVittie model and a specific LTB model. Our work reveals interesting features of the measured curvature in the McVittie model that involves the linear superposition of the curvature due to the inhomogeneity, the curvature of the background FLRW universe and a coupling term.  The contribution of the inhomogeneity to the measured spacetime curvature in the more standard LTB model is of a different nature. Despite the similarities between these inhomogeneous cosmological models, their tidal dynamics are very different. The tidal dynamics in our LTB model has the same character as in the Einstein-de Sitter model. Furthermore, the McVittie model forces the background FLRW Hubble flow to become accelerated in order to prevent accretion onto the source of the inhomogeneity. This circumstance adds further difficulty to the possibility of comparison with observational data. In addition, it is not clear in general how to connect the motion of particles within a Fermi normal coordinate system that is located inside a distant galaxy in some cosmological model with local observational data regarding peculiar motions. The resolutions of these difficulties constitute tasks for the future. 

%%%%%%%%%%%%%
\section*{ACKNOWLEDGMENTS}
We thank Javad Tabatabaei for helpful discussions.
SB is partially supported by the Abdus Salam International Center for Theoretical Physics (ICTP) under the regular associateship scheme.
Moreover, MM and SB are partially supported by the Sharif University of Technology Office of Vice President for Research under Grant No. G4010204. 
%%%%%%%%%%%%%%%

\appendix

\section{Connection coefficients}

Let us start with the flat McVittie metric~\eqref{J1} in spherical polar coordinates $(t, \rho, \theta, \phi)$. It is necessary to compute  $\Gamma^{\mu}_{\nu \rho} =  \Gamma^{\mu}_{\rho \nu}$ in terms of $\rho$ and $a(t)$. The nonzero connection coefficients can be calculated using
\begin{equation}\label{A1}
 \Gamma^{0}_{0 0} = \rho H \, \Gamma^{0}_{0 1}\,, \qquad \Gamma^{0}_{0 1} = \frac{4 M a}{PQ}\,,
\end{equation}  
\begin{equation}\label{A2}
\Gamma^{0}_{2 2} = \rho^2\, \Gamma^{0}_{1 1} = \frac{H Q^5}{16 \rho^2 a^2P}\,,\qquad  \Gamma^{0}_{3 3} = \sin^2 \theta \, \Gamma^{0}_{2 2}\,, 
\end{equation} 
\begin{equation}\label{A3}
\Gamma^{1}_{0 0} = \frac{64\,M \rho^4 a^3P}{Q^7}\,, \qquad \Gamma^{1}_{1 1}= -\frac{2\,M}{\rho \,Q}\,, 
\end{equation} 
\begin{equation}\label{A4}
\Gamma^{1}_{0 1} = H\,\frac{P}{Q}\,, \qquad \Gamma^{1}_{2 2}= - \rho\, \frac{P}{Q}\,, \qquad \Gamma^{1}_{3 3}= \sin^2\theta\, \Gamma^{1}_{2 2}\,, 
\end{equation} 
\begin{equation}\label{A5}
\Gamma^{2}_{0 2} = \Gamma^{3}_{0 3}  = H\,\frac{P}{Q}\,, \qquad \Gamma^{2}_{1 2}= \Gamma^{3}_{1 3} = \frac{1}{\rho}\, \frac{P}{Q}\,, 
\end{equation} 
\begin{equation}\label{A6}
\Gamma^{2}_{3 3} = - \sin\theta \cos \theta\,,\qquad \Gamma^{3}_{2 3}  = \cot \theta\,. 
\end{equation} 
Other nonzero components can be obtained from the symmetry of the Christoffel symbols. 

\section{Explicit Form of Eqs.~\eqref{L5}--\eqref{L7}}

The long versions of Eqs.~\eqref{L5}--\eqref{L7} for $\gamma = 0$ are given below for the sake of completeness. We assume $\mathcal{E}_2 = \mathcal{E}_3$ and $\mathcal{S}_2 = \mathcal{S}_3$.
\begin{align}\label{B1}
 \nonumber  \frac{dV^1}{dT} {} & +  \mathcal{E}_1\,[1-2(V^1)^2]\,X^1  - 2\,\mathcal{E}_2 \,V^1\,(X^2\,V^2+X^3\,V^3) \\  
& +\frac{2}{3}\,\mathcal{S}_2\,\{ X^1\, [(V^2)^2 + (V^3)^2] - X^2\, V^1\,V^2 - X^3 \,V^1\,V^3\}   = 0\,, 
\end{align}
\begin{align}\label{B2}
\nonumber  \frac{dV^2}{dT} {} & +  \mathcal{E}_2\,[1-2(V^2)^2]\,X^2  - 2\, V^2\,(\mathcal{E}_1\,X^1\,V^1+\mathcal{E}_2\,X^3\,V^3) \\   
& +\frac{2}{3}\,\{- \mathcal{S}_2\,X^1\, V^1\,V^2 + X^2\, [\mathcal{S}_1\,(V^3)^2 + \mathcal{S}_2\,(V^1)^2]  - \mathcal{S}_1\,X^3 \,V^2\,V^3\}   = 0\,, 
\end{align}
\begin{align}\label{B3}
\nonumber  \frac{dV^3}{dT} {} & +  \mathcal{E}_2\,[1-2(V^3)^2]\,X^3  - 2\, V^3\,(\mathcal{E}_1\,X^1\,V^1+\mathcal{E}_2\,X^2\,V^2) \\   
& +\frac{2}{3}\,\{- \mathcal{S}_2\,X^1\, V^1\,V^3 - \mathcal{S}_1\,X^2 \,V^2\,V^3 + X^3\, [\mathcal{S}_1\,(V^2)^2 + \mathcal{S}_2\,(V^1)^2] \}   = 0\,.
\end{align}


\begin{thebibliography}{99}

%\cite{Bini:2014esa}
\bibitem{Bini:2014esa}
D.~Bini and B.~Mashhoon,
``Peculiar velocities in dynamic spacetimes",
Phys. Rev. D \textbf{90}, no.2, 024030 (2014).
%doi:10.1103/PhysRevD.90.024030
[arXiv:1405.4430 [gr-qc]]
%10 citations counted in INSPIRE as of 14 Jul 2024

%\cite{Chicone:2010xr}
\bibitem{Chicone:2010xr}
C.~Chicone, B.~Mashhoon and K.~Rosquist,
``Cosmic Jets",
Phys. Lett. A \textbf{375}, 1427-1430 (2011).
%doi:10.1016/j.physleta.2011.02.036
[arXiv:1011.3477 [gr-qc]]
%15 citations counted in INSPIRE as of 14 Jul 2024


%\cite{Chicone:2011ie}
\bibitem{Chicone:2011ie}
C.~Chicone, B.~Mashhoon and K.~Rosquist,
``Double-Kasner Spacetime: Peculiar Velocities and Cosmic Jets",
Phys. Rev. D \textbf{83}, 124029 (2011).
%doi:10.1103/PhysRevD.83.124029
[arXiv:1104.5058 [gr-qc]]
%12 citations counted in INSPIRE as of 14 Jul 2024


%\cite{Dodelson:2003ft}
\bibitem{Dodelson:2003ft}
S.~Dodelson,
{\it{Modern Cosmology}} (Academic Press, 2003).
%ISBN 978-0-12-219141-1
%419 citations counted in INSPIRE as of 04 Jun 2024


%\cite{Weinberg:2008zzc}
\bibitem{Weinberg:2008zzc}
S.~Weinberg, {\it{Cosmology}} (Oxford University Press, 2008).
%525 citations counted in INSPIRE as of 04 Jun 2024

%\cite{Amendola:2015ksp}
\bibitem{Amendola:2015ksp}
L.~Amendola and S.~Tsujikawa,
\emph{Dark Energy: Theory and Observations}
(Cambridge University Press, 2015)
%ISBN 978-1-107-45398-2
%16 citations counted in INSPIRE as of 16 Nov 2024



%\cite{Baumann:2022mni}
\bibitem{Baumann:2022mni}
D.~Baumann, {\it{Cosmology}} (Cambridge University Press, 2022).
%ISBN 978-1-108-93709-2, 978-1-108-83807-8
%doi:10.1017/9781108937092
%46 citations counted in INSPIRE as of 04 Jun 2024


%\cite{Mohayaee:2020wxf}
\bibitem{Mohayaee:2020wxf}
R.~Mohayaee, M.~Rameez and S.~Sarkar,
``Cosmological Inference from within the Peculiar Local Universe",
Universe \textbf{10}, no.5, 209 (2024).
%doi:10.3390/universe10050209
[arXiv:2003.10420 [astro-ph.CO]]
%23 citations counted in INSPIRE as of 06 Sep 2024

\bibitem{Immer}
K.  Immer, J.  Li, L. H.  Quiroga-Nunez, M. J.  Reid, B.  Zhang, L.  Moscadelli and K. L. J.  Rygl,
``Anomalous peculiar motions of high-mass young stars in the Scutum spiral arm",
Astron. \& Astrophys. \textbf{632},  A123 (2019).
[arXiv:1911.06806 [astro-ph.GA]]
%DOI: 10.1051/0004-6361/201834208

\bibitem{Zin}
I. A.  Zinchenko, L. S.  Pilyugin, F.  Sakhibov, E. K.  Grebel, A.  Just, P.  Berczik, Y. A.  Nefedyev, J. M.  Vilchez and V. M.  Shulga,
``Peculiar motions of the gas at the centre of the barred galaxy UGC 4056",
Astron. \& Astrophys. \textbf{628},  A55 (2019).
[arXiv:1907.07998 [astro-ph.GA]]
%DOI: 10.1051/0004-6361/201935897

\bibitem{Pesce}
D. W. Pesce, J. A. Braatz, J. J. Condon and J. E. Greene,
``Measuring Supermassive Black Hole Peculiar Motion Using H2O Megamasers",
Astrophys. J. \textbf{863}, 149 (2018).
[arXiv:1807.04598 [astro-ph.GA]]
%DOI 10.3847/1538-4357/aad3c2


%\cite{Giahi-Saravani:2012rou}
\bibitem{Giahi-Saravani:2012rou}
A.~Giahi-Saravani and B.~M.~Schaefer,
``Evolution of intrinsic ellipticity correlations due to peculiar motion",
Mon. Not. Roy. Astron. Soc. \textbf{428}, 1312-1320 (2013).
%doi:10.1093/mnras/sts110
[arXiv:1202.1196 [astro-ph.CO]]
%10 citations counted in INSPIRE as of 06 Sep 2024

%\cite{Baba:2009ep}
\bibitem{Baba:2009ep}
J.~Baba, Y.~Asaki, J.~Makino, M.~Miyoshi, T.~R.~Saitoh and K.~Wada,
``The origin of large peculiar motions of star-forming regions and spiral structures of our Galaxy",
Astrophys. J. \textbf{706}, 471-481 (2009).
%doi:10.1088/0004-637X/706/1/471
[arXiv:0904.4305 [astro-ph.GA]]
%28 citations counted in INSPIRE as of 06 Sep 2024


\bibitem{Einstein} 
A. Einstein, \emph{The Meaning of Relativity}
(Princeton University Press, Princeton, NJ, USA, 1955). 

%\cite{Szekeres:1974ct}
\bibitem{Szekeres:1974ct}
P.~Szekeres,
``A Class of Inhomogeneous Cosmological Models",
Commun. Math. Phys. \textbf{41}, 55-64 (1975).
%doi:10.1007/BF01608547
%324 citations counted in INSPIRE as of 19 Jul 2024


%\cite{Lemaitre:1933gd}
\bibitem{Lemaitre:1933gd}
G.~Lema\^itre,
``L'univers en expansion",
Annales Soc. Sci. Bruxelles A \textbf{53}, 51-85 (1933).
%doi:10.1023/A:1018855621348
[Reprinted with English Translation: ``The expanding universe", Gen. Relativ. Gravit. \textbf{29}, 641-680 (1997).]
%764 citations counted in INSPIRE as of 18 Jul 2024

%\cite{Tolman:1934za}
\bibitem{Tolman:1934za}
R.~C.~Tolman,
``Effect of imhomogeneity on cosmological models",
Proc. Nat. Acad. Sci. \textbf{20}, 169-176 (1934).
%doi:10.1073/pnas.20.3.169
[Reprinted: Gen. Relativ. Gravit. \textbf{29}, 935 (1997).]
%857 citations counted in INSPIRE as of 18 Jul 2024

%\cite{Bondi:1947fta}
\bibitem{Bondi:1947fta}
H.~Bondi,
``Spherically symmetrical models in general relativity",
Mon. Not. Roy. Astron. Soc. \textbf{107}, 410-425 (1947).
%doi:10.1093/mnras/107.5-6.410
[Reprinted: Gen. Relativ. Gravit. \textbf{31}, 1777 (1999).]
%744 citations counted in INSPIRE as of 18 Jul 2024



%\cite{Krasinski:1997yxj}
\bibitem{Krasinski:1997yxj}
A.~Krasinski,
\emph{Inhomogeneous Cosmological Models}
(Cambridge University Press, 2006).
%ISBN 978-0-511-88754-3, 978-0-521-48180-9, 978-0-521-03017-5
%37 citations counted in INSPIRE as of 30 May 2024



%\cite{Plebanski:2006sd}
\bibitem{Plebanski:2006sd}
J.~Plebanski and A.~Krasinski,
\emph{An introduction to general relativity and cosmology}, 2nd edn
(Cambridge University Press, 2024).
%41 citations counted in INSPIRE as of 30 May 2024


%\cite{Bolejko:2009pvd}
\bibitem{Bolejko:2009pvd}
K.~Bolejko, A.~Krasinski, C.~Hellaby and M.~N.~Celerier,
\emph{Structures in the Universe by Exact Methods}
(Cambridge University Press, 2009).
%doi:10.1017/CBO9780511657405
%8 citations counted in INSPIRE as of 30 May 2024


%\cite{Ribeiro:1992iwc}
\bibitem{Ribeiro:1992iwc}
M.~B.~Ribeiro,
``On Modelling a Relativistic Hierarchical (Fractal) Cosmology by Tolman's Spacetime. I. Theory",
Astrophys. J. \textbf{388}, 1-8 (1992).
%doi:10.1086/171123
[arXiv:0807.0866 [astro-ph]]
%58 citations counted in INSPIRE as of 08 Nov 2024

%\cite{Marra:2007pm}
\bibitem{Marra:2007pm}
V.~Marra, E.~W.~Kolb, S.~Matarrese and A.~Riotto,
``On cosmological observables in a swiss-cheese universe",
Phys. Rev. D \textbf{76}, 123004 (2007).
%doi:10.1103/PhysRevD.76.123004
[arXiv:0708.3622 [astro-ph]]
%187 citations counted in INSPIRE as of 08 Nov 2024

%\cite{Marra:2007gc}
\bibitem{Marra:2007gc}
V.~Marra, E.~W.~Kolb and S.~Matarrese,
``Light-cone averages in a swiss-cheese Universe",
Phys. Rev. D \textbf{77}, 023003 (2008).
%doi:10.1103/PhysRevD.77.023003
[arXiv:0710.5505 [astro-ph]]
%126 citations counted in INSPIRE as of 08 Nov 2024


%\cite{Duffy:2010bu}
\bibitem{Duffy:2010bu}
E.~M.~Duffy and B.~C.~Nolan,
``Odd Parity Perturbations of the Self-Similar LTB Spacetime",
Classical Quantum Gravity \textbf{28}, 105020 (2011).
%doi:10.1088/0264-9381/28/10/105020
[arXiv:1012.2766 [gr-qc]]
%8 citations counted in INSPIRE as of 08 Nov 2024

%\cite{Cosmai:2018nvx}
\bibitem{Cosmai:2018nvx}
L.~Cosmai, G.~Fanizza, F.~Sylos Labini, L.~Pietronero and L.~Tedesco,
``Fractal universe and cosmic acceleration in a Lema\^\i{}tre\textendash{}Tolman\textendash{}Bondi scenario",
Classical Quantum Gravity \textbf{36}, no.4, 045007 (2019).
%doi:10.1088/1361-6382/aae8f7
[arXiv:1810.06318 [astro-ph.CO]]
%20 citations counted in INSPIRE as of 08 Nov 2024


%\cite{McVittie:1933zz}
\bibitem{McVittie:1933zz}
G.~C.~McVittie,
``The mass-particle in an expanding universe",
Mon. Not. Roy. Astron. Soc. \textbf{93}, 325-339 (1933).
%\url{https://doi:10.1093/mnras/93.5.325}
%374 citations counted in INSPIRE as of 12 Oct 2023

%\cite{Hahn:2014lca}
\bibitem{Hahn:2014lca}
O.~Hahn, R.~E.~Angulo and T.~Abel,
``The Properties of Cosmic Velocity Fields,''
Mon. Not. Roy. Astron. Soc. \textbf{454}, no.4, 3920-3937 (2015).
%doi:10.1093/mnras/stv2179
[arXiv:1404.2280 [astro-ph.CO]]
%105 citations counted in INSPIRE as of 16 Nov 2024
 
%\cite{Watkins:2008hf}
\bibitem{Watkins:2008hf}
R.~Watkins, H.~A.~Feldman and M.~J.~Hudson,
``Consistently Large Cosmic Flows on Scales of 100 Mpc/h: a Challenge for the Standard LCDM Cosmology",
Mon. Not. Roy. Astron. Soc. \textbf{392}, 743-756 (2009).
%doi:10.1111/j.1365-2966.2008.14089.x
[arXiv:0809.4041 [astro-ph]]
%299 citations counted in INSPIRE as of 16 Nov 2024
 
 
%\cite{Davis:2010jq}
\bibitem{Davis:2010jq}
T.~M.~Davis, L.~Hui, J.~A.~Frieman, T.~Haugbolle, R.~Kessler, B.~Sinclair, J.~Sollerman, B.~Bassett, J.~Marriner and E.~Mortsell, \textit{et al.}
``The Effect of Peculiar Velocities on Supernova Cosmology",
Astrophys. J. \textbf{741}, 67 (2011).
%doi:10.1088/0004-637X/741/1/67
[arXiv:1012.2912 [astro-ph.CO]]
%122 citations counted in INSPIRE as of 16 Nov 2024


\bibitem{Synge}
J.~L.~Synge, 
\emph{Relativity: The General Theory} (North-Holland, Amsterdam, 1971).  

\bibitem{mash77}
B. Mashhoon,
``Tidal radiation'',
Astrophys. J.\ {\bf 216}, 591-609 (1977).

%\cite{Chicone:2002kb}
\bibitem{Chicone:2002kb}
C.~Chicone and B.~Mashhoon,
``The generalized Jacobi equation",
Classical Quantum Gravity \textbf{19}, 4231-4248 (2002).
%doi:10.1088/0264-9381/19/16/301
[arXiv:gr-qc/0203073 [gr-qc]]
%52 citations counted in INSPIRE as of 09 May 2024

%\cite{Chicone:2005vn}
\bibitem{Chicone:2005vn}
C.~Chicone and B.~Mashhoon,
``Explicit Fermi coordinates and tidal dynamics in de Sitter and G\"odel spacetimes",
Phys. Rev. D \textbf{74}, 064019 (2006).
%doi:10.1103/PhysRevD.74.064019
[arXiv:gr-qc/0511129 [gr-qc]]
%61 citations counted in INSPIRE as of 24 May 2024

\bibitem{Mashhoon}
B. Mashhoon, 
``Is the Universe Homogeneous on a Large Scale?", 
in: \emph{The Big Bang and Georges Lema\^itre}, edited by A. Berger (Reidel, Dordrecht, 1984), pp. 75-81. 





%\cite{Cooperstock:1998ny}
\bibitem{Cooperstock:1998ny}
F.~I.~Cooperstock, V.~Faraoni and D.~N.~Vollick,
``The Influence of the cosmological expansion on local systems",
Astrophys. J. \textbf{503}, 61 (1998).
%doi:10.1086/305956
[arXiv:astro-ph/9803097 [astro-ph]]
%89 citations counted in INSPIRE as of 31 May 2024


%\cite{Mashhoon:2007qm}
\bibitem{Mashhoon:2007qm}
B.~Mashhoon, N.~Mobed and D.~Singh,
``Tidal dynamics in cosmological spacetimes",
Classical Quantum Gravity \textbf{24}, 5031-5046 (2007).
%doi:10.1088/0264-9381/24/20/008
[arXiv:0705.1312 [gr-qc]]
%36 citations counted in INSPIRE as of 25 May 2024


%\cite{Nandra:2011ug}
\bibitem{Nandra:2011ug}
R.~Nandra, A.~N.~Lasenby and M.~P.~Hobson,
``The effect of a massive object on an expanding universe",
Mon. Not. Roy. Astron. Soc. \textbf{422}, 2931-2944 (2012).
%doi:10.1111/j.1365-2966.2012.20618.x
[arXiv:1104.4447 [gr-qc]]
%72 citations counted in INSPIRE as of 30 May 2024

%\cite{Nandra:2011ui}
\bibitem{Nandra:2011ui}
R.~Nandra, A.~N.~Lasenby and M.~P.~Hobson,
``The effect of an expanding universe on massive objects",
Mon. Not. Roy. Astron. Soc. \textbf{422}, 2945-2959 (2012).
%doi:10.1111/j.1365-2966.2012.20617.x
[arXiv:1104.4458 [gr-qc]]
%61 citations counted in INSPIRE as of 30 May 2024

%\cite{Nandra:2013jga}
\bibitem{Nandra:2013jga}
R.~Nandra, A.~Lasenby and M.~Hobson,
``Dynamics of a spherical object of uniform density in an expanding universe",
Phys. Rev. D \textbf{88}, no.4, 044041 (2013).
%doi:10.1103/PhysRevD.88.044041
[arXiv:1307.0526 [astro-ph.CO]]
%3 citations counted in INSPIRE as of 30 May 2024

%\cite{Faraoni:2007es}
\bibitem{Faraoni:2007es}
V.~Faraoni and A.~Jacques,
``Cosmological expansion and local physics",
Phys. Rev. D \textbf{76}, 063510 (2007).
%doi:10.1103/PhysRevD.76.063510
[arXiv:0707.1350 [gr-qc]]
%152 citations counted in INSPIRE as of 31 May 2024


%\cite{Kopeikin:2012by}
\bibitem{Kopeikin:2012by}
S.~Kopeikin,
``Celestial Ephemerides in an Expanding Universe",
Phys. Rev. D \textbf{86}, 064004 (2012).
%doi:10.1103/PhysRevD.86.064004
[arXiv:1207.3873 [gr-qc]]
%33 citations counted in INSPIRE as of 31 May 2024


%\cite{Kopeikin:2013am}
\bibitem{Kopeikin:2013am}
S.~M.~Kopeikin and A.~N.~Petrov,
``Post-Newtonian Celestial Dynamics in Cosmology: Field Equations",
Phys. Rev. D \textbf{87}, no.4, 044029 (2013).
%doi:10.1103/PhysRevD.87.044029
[arXiv:1301.5706 [gr-qc]]
%16 citations counted in INSPIRE as of 31 May 2024


%\cite{Iorio:2012wva}
\bibitem{Iorio:2012wva}
L.~Iorio,
``Local cosmological effects of order H in the orbital motion of a binary system?",
Mon. Not. Roy. Astron. Soc. \textbf{429}, 915-922 (2013).
%doi:10.1093/mnras/sts396
[arXiv:1208.1523 [gr-qc]]
%13 citations counted in INSPIRE as of 31 May 2024

%\cite{Spengler:2021vxy}
\bibitem{Spengler:2021vxy}
F.~Spengler, A.~Belenchia, D.~R\"atzel and D.~Braun,
``Influence of cosmological expansion in local experiments",
Classical Quantum Gravity \textbf{39}, no.5, 055005 (2022).
%doi:10.1088/1361-6382/ac4954
[arXiv:2109.03280 [gr-qc]]
%1 citations counted in INSPIRE as of 31 May 2024

%\cite{Ferraris:1996ey}
\bibitem{Ferraris:1996ey}
M.~Ferraris, M.~Francaviglia and A.~Spallicci,
``Associated radius, energy and pressure of McVittie's metric, in its astrophysical application",
Nuovo Cim. B \textbf{111}, 1031-1036 (1996).
%doi:10.1007/BF02743299
%34 citations counted in INSPIRE as of 08 Nov 2024


\bibitem{Bonnor}
W. B. Bonnor, 
`` Local Dynamics and the Expansion of the Universe", 
Gen. Relativ. Gravit. \textbf{32}, 1005-1007 (2000).
%DOI: 10.1023/A:1001961325184 

%\cite{Sakai:1999xx}
\bibitem{Sakai:1999xx}
N.~Sakai and P.~Haines,
``Peculiar velocities of nonlinear structure: Voids in McVittie space-time",
Astrophys. J. \textbf{536}, 515 (2000).
%doi:10.1086/308965
[arXiv:astro-ph/9909183 [astro-ph]]
%6 citations counted in INSPIRE as of 08 Nov 2024


%\cite{Kaloper:2010ec}
\bibitem{Kaloper:2010ec}
N.~Kaloper, M.~Kleban and D.~Martin,
``McVittie's Legacy: Black Holes in an Expanding Universe",
Phys. Rev. D \textbf{81}, 104044 (2010).
%\url{https://doi:10.1103/PhysRevD.81.104044},
[arXiv:1003.4777 [hep-th]]
%136 citations counted in INSPIRE as of 12 Oct 2023

%\cite{Lake:2011ni}
\bibitem{Lake:2011ni}
K.~Lake and M.~Abdelqader,
``More on McVittie's Legacy: A Schwarzschild - de Sitter black and white hole embedded in an asymptotically $\Lambda$CDM cosmology",
Phys. Rev. D \textbf{84}, 044045 (2011).
%\url{https://doi:10.1103/PhysRevD.84.044045},
[arXiv:1106.3666 [gr-qc]]
%78 citations counted in INSPIRE as of 12 Oct 2023

%\cite{Nolan:2014maa}
\bibitem{Nolan:2014maa}
B.~C.~Nolan,
``Particle and photon orbits in McVittie spacetimes",
Classical Quantum Gravity \textbf{31}, no.23, 235008 (2014).
%doi:10.1088/0264-9381/31/23/235008
[arXiv:1408.0044 [gr-qc]]
%34 citations counted in INSPIRE as of 30 May 2024

%\cite{Nolan:2017rtj}
\bibitem{Nolan:2017rtj}
B.~C.~Nolan,
``Local properties and global structure of McVittie spacetimes with non-flat Friedmann\textendash{}Lema\^\i{}tre\textendash{}Robertson\textendash{}Walker backgrounds",
Classical Quantum Gravity \textbf{34}, no.22, 225002 (2017).
%\url{https://doi:10.1088/1361-6382/aa903c},
[arXiv:1707.07612 [gr-qc]]
%8 citations counted in INSPIRE as of 12 Oct 2023

%\cite{Perlick:2018iye}
\bibitem{Perlick:2018iye}
V.~Perlick, O.~Y.~Tsupko and G.~S.~Bisnovatyi-Kogan,
``Black hole shadow in an expanding universe with a cosmological constant",
Phys. Rev. D \textbf{97}, no.10, 104062 (2018).
%\url{https://doi:10.1103/PhysRevD.97.104062},
[arXiv:1804.04898 [gr-qc]]
%134 citations counted in INSPIRE as of 12 Oct 2023

%\cite{Faraoni:2018xwo}
\bibitem{Faraoni:2018xwo}
V.~Faraoni,
``Embedding black holes and other inhomogeneities in the universe in various theories of gravity: a short review",
Universe \textbf{4}, no.10, 109 (2018).
%\url{https://doi:10.3390/universe4100109},
[arXiv:1810.04667 [gr-qc]]
%27 citations counted in INSPIRE as of 12 Oct 2023

%\cite{Rothman:2018haq}
\bibitem{Rothman:2018haq}
T.~Rothman, M.~Campbell, R.~Goswami and G.~F.~R.~Ellis,
``Direct Detection of Universal Expansion by Holonomy in the McVittie Spacetime",
Phys. Rev. D \textbf{99}, no.2, 024033 (2019).
%doi:10.1103/PhysRevD.99.024033
[arXiv:1808.02963 [gr-qc]]
%0 citations counted in INSPIRE as of 08 Nov 2024


%\cite{Gaur:2023hmk}
\bibitem{Gaur:2023hmk}
R.~Gaur and M.~Visser,
``Black holes embedded in FLRW cosmologies",
Phys. Rev. D \textbf{110}, no.4, 043529 (2024).
%doi:10.1103/PhysRevD.110.043529
[arXiv:2308.07374 [gr-qc]]
%14 citations counted in INSPIRE as of 08 Nov 2024


\bibitem{Kottler}
F. Kottler, 
``\"Uber die physikalischen Grundlagen der Einsteinschen Gravitationstheorie",
Ann. Phys. (Leipzig) \textbf{56}, 410-462 (1918). 
% https://doi.org/10.1002/andp.19183611402


%\cite{Misner:1964je}
\bibitem{Misner:1964je}
C.~W.~Misner and D.~H.~Sharp,
``Relativistic equations for adiabatic, spherically symmetric gravitational collapse",
Phys. Rev. \textbf{136}, B571-B576 (1964). 
%\url{https://doi:10.1103/PhysRev.136.B571}
%1047 citations counted in INSPIRE as of 02 Feb 2024

\bibitem{HeMi}
W. C. Hernandez, Jr. and C. W. Misner,
``Observer time as a coordinate in relativistic spherical hydrodynamics",
Astrophys. J. \textbf{143}, 452-464 (1966).

\bibitem{CaMc}
M. E. Cahill and G. C. McVittie,
``Spherical Symmetry and Mass-Energy in General Relativity. I. General Theory", 
J. Math. Phys. \textbf{11}, 1382-1391 (1970).
%https://doi.org/10.1063/1.1665273

\bibitem{GlaMa}
E. N. Glass and B. Mashhoon, 
``On a spherical star system with a collapsed core", 
Astrophys. J. \textbf{205}, 570-577 (1976).


%\cite{Mashhoon:1979tt}
\bibitem{Mashhoon:1979tt}
B.~Mashhoon and M.~H.~Partovi,
``Gravitational Collapse of a Charged Fluid Sphere",
Phys. Rev. D \textbf{20}, 2455 (1979).
%\url{https://doi:10.1103/PhysRevD.20.2455}
%30 citations counted in INSPIRE as of 09 Oct 2023

%\cite{Mashhoon:2020tha}
\bibitem{Mashhoon:2020tha}
B.~Mashhoon,
``Critical Tidal Currents in General Relativity",
Universe \textbf{6}, no.8, 104 (2020).
%doi:10.3390/universe6080104
[arXiv:2007.12023 [gr-qc]]
%2 citations counted in INSPIRE as of 11 Sep 2024

%\cite{Krasinski:2010pfm}
\bibitem{Krasinski:2010pfm}
A.~Krasinski, C.~Hellaby, K.~Bolejko and M.~N.~Celerier,
``Imitating accelerated expansion of the Universe by matter inhomogeneities: Corrections of some misunderstandings",
Gen. Relativ. Gravit. \textbf{42}, 2453-2475 (2010).
%doi:10.1007/s10714-010-0993-5
[arXiv:0903.4070 [gr-qc]]
%46 citations counted in INSPIRE as of 09 Oct 2024


%\cite{Pajer:2013ana}
\bibitem{Pajer:2013ana}
E.~Pajer, F.~Schmidt and M.~Zaldarriaga,
``The Observed Squeezed Limit of Cosmological Three-Point Functions",
Phys. Rev. D \textbf{88}, no.8, 083502 (2013).
%doi:10.1103/PhysRevD.88.083502
[arXiv:1305.0824 [astro-ph.CO]]
%204 citations counted in INSPIRE as of 21 Sep 2024

%\cite{Dai:2015rda}
\bibitem{Dai:2015rda}
L.~Dai, E.~Pajer and F.~Schmidt,
``Conformal Fermi Coordinates",
JCAP \textbf{11}, 043 (2015).
%doi:10.1088/1475-7516/2015/11/043
[arXiv:1502.02011 [gr-qc]]
%62 citations counted in INSPIRE as of 21 Sep 2024



\end{thebibliography}
\end{document}